\documentclass[aoas,MSNbibl,nameyear,dvips]{arximspdf}
\usepackage{dcolumn}
\usepackage{graphicx}

\doi{10.1214/14-AOAS748} % kopijuoti is laisko
\volume{8}
\issue{4}
\pubyear{2014}
\firstpage{2175}
\lastpage{2202}
\docsubty{FLA}

\makeatletter
\newtheorem{lemma}{Lemma}
\newcolumntype{d}[1]{D{.}{.}{#1}}

\newcommand{\bbY}{\widetilde{\mathbf{Y}}}
\newcommand{\bhK}{\hat{\mathbf{K}}}
\newcommand{\bhA}{\hat{\mathbf{A}}}
\newcommand{\bY}{{\mathbf{Y}}}
\newcommand{\bI}{{\mathbf{I}}}
\newcommand{\bK}{{\mathbf{K}}}
\newcommand{\bU}{{\mathbf{U}}}
\newcommand{\bV}{{\mathbf{V}}}
\newcommand{\bA}{{\mathbf{A}}}
\newcommand{\bB}{{\mathbf{B}}}
\newcommand{\bZ}{{\mathbf{Z}}}
\newcommand{\bZero}{{\mathbf{0}}}
\newcommand{\bF}{{\mathbf{F}}}
\newcommand{\bH}{{\mathbf{H}}}
\newcommand{\bD}{{\mathbf{D}}}
\newcommand{\bS}{{\mathbf{S}}}
\newcommand{\bff}{{\mathbf{f}}}
\newcommand{\mcV}{\mathcal{V}}
\newcommand{\bLambda}{\bolds{\Lambda}}
\newcommand{\bPhi}{\bolds{\Phi}}
\newcommand{\bomega}{\bolds{\omega}}
\newcommand{\bxi}{\bolds{\xi}}
\newcommand{\bzeta}{\bolds{\zeta}}
\newcommand{\hbeta}{\hat{\bolds{\eta}}}
\makeatother

\begin{document}
\begin{frontmatter}

\title{Longitudinal high-dimensional principal components
analysis with application to diffusion tensor imaging of multiple
sclerosis}
\runtitle{HD-LFPCA}

\begin{aug}
\author[A]{\fnms{Vadim}~\snm{Zipunnikov}\corref{}\thanksref{t1,TT1}\ead[label=e1]{vzipunn1@jhu.edu}},
\author[B]{\fnms{Sonja}~\snm{Greven}\thanksref{t2,TT2}},
\author[C]{\fnms{Haochang} \snm{Shou}\thanksref{t4}},
\author[A]{\fnms{Brian~S.} \snm{Caffo}\thanksref{t1,TT1}},
\author[D]{\fnms{Daniel~S.}~\snm{Reich}\thanksref{t3,TT3}}\break 
\and
\author[A]{\fnms{Ciprian~M.}~\snm{Crainiceanu}\thanksref{t1,TT1}}
\runauthor{V. Zipunnikov et al.}
\affiliation{Johns Hopkins University\thanksmark{t1},
Ludwig-Maximilians-Universit\"at M\"unchen\thanksmark{t2},
University of Pennsylvania\thanksmark{t4}
and National Institutes of Health\thanksmark{t3}}

\thankstext{TT1}{Supported by Grant R01NS060910 from the
National Institute of Neurological Disorders and Stroke and by
Award Number EB012547 from the NIH National Institute of Biomedical
Imaging and Bioengineering (NIBIB).}

\thankstext{TT2}{Supported by the German Research
Foundation through the Emmy Noether Programme, Grant GR 3793/1-1.}

\thankstext{TT3}{Supported by the Intramural
Research Program of the National Institute of Neurological
Disorders and Stroke.}

\address[A]{V. Zipunnikov\\
B. S. Caffo\\
C. M. Crainiceanu\\
Department of Biostatistics\\
Johns Hopkins University\\
Baltimore, Maryland 21205-2179\\
USA\\
\printead{e1}} %adresu isvedimo komanda gale!
\address[B]{S. Greven\\
Department of Statistics\\
Ludwig-Maximilians-University\hspace*{37pt}\\
80539 Munich\\
Germany}
\address[C]{H. Shou\\
Department of Biostatistics\\
\quad and Epidemiology\\
University of Pennsylvania\\
Philadelphia, Pennsylvania 19104-6021\\
USA}
\address[D]{D. S. Reich\\
National Institute \\
\quad of Neurological Disorders and Stroke\\
National Institutes of Health\\
Bethesda, Maryland 20824\\
USA}
\end{aug}

% HISTORY:
\received{\smonth{8} \syear{2013}}
\revised{\smonth{3} \syear{2014}}

% ABSTRACT
%
\begin{abstract}
We develop a flexible framework for modeling high-dimensional imaging
data observed longitudinally. The approach decomposes the observed
variability of repeatedly measured high-dimensional observations into
three additive components: a subject-specific imaging random intercept
that quantifies the cross-sectional variability, a\break subject-specific
imaging slope that quantifies the dynamic irreversible deformation over
multiple realizations, and a~subject-visit-specific imaging deviation
that quantifies exchangeable effects between visits. The proposed
method is very fast, scalable to studies including
ultrahigh-dimensional data, and can easily be adapted to and executed
on modest computing infrastructures. The method is applied to the
longitudinal analysis of diffusion tensor imaging (DTI) data of the
corpus callosum of multiple sclerosis (MS) subjects. The study includes
$176$ subjects observed at $466$ visits. For each subject and visit the
study contains a registered DTI scan of the corpus callosum at roughly
30,000 voxels.
\end{abstract}

% KEYWORDS
% Pirmas kwd is didziosios raides
%
\end{frontmatter}

%s1 #&#
\section{Introduction}\label{sec:Introduction}
% General stuff; how many studies
An increasing number of longitudinal studies routinely acquire
high-dimensional data, such as brain images or gene expression, at
multiple visits. This led to increased interest in generalizing
standard models designed for longitudinal data analysis to the case
when the observed data are massively multivariate. In this paper we
propose to generalize the random intercept random slope mixed effects
model to the case when instead of a scalar, one measures a
high-dimensional object, such as a brain image. {The proposed
methods can be applied to longitudinal studies that include
high-dimensional imaging observations without missing data that can be
unfolded into a long vector}.

% Describe the data
This paper is motivated by a study of multiple sclerosis (MS) patients
[\citet{Reich:etal:2010}].
Multiple sclerosis is a degenerative disease of the central nervous
system. A hallmark of MS is damage to and degeneration of the myelin
sheaths that surround and insulate nerve fibers in the brain. Such
damage results in sclerotic plaques that distort the flow of electrical
impulses along the nerves to different parts of the body [\citet
{Raine:etal:2008}]. MS also affects the neurons themselves and is
associated with accelerated brain atrophy.

Our data are derived from a natural history study of $176$ MS cases
selected from a population with a wide spectrum of disease severity.
Subjects were scanned over a 5-year period up to $10$ times per
subject, for a total of 466 scans. The scans have been aligned
(registered) using a $12$ degrees of freedom transformation which
accounts for rotation, translation, scaling, and shearing, but not for
nonlinear deformation. In this study we focus on fractional anisotropy
(FA), a useful voxel-level summary of diffusion tensor imaging (DTI), a
type of structural Magnetic Resonance Imaging (MRI). FA is viewed as a
measure of tissue integrity and is thought to be sensitive both to axon
fiber density and myelination in white matter. {It is measured
on a scale between zero (isotropic diffusion characteristic of
fluid-filled cavities) and one (anisotropic diffusion, characteristic
of highly ordered white matter fiber bundles) [\citet{Mori:2007}].}

%f1 #&#
\begin{figure}

\includegraphics{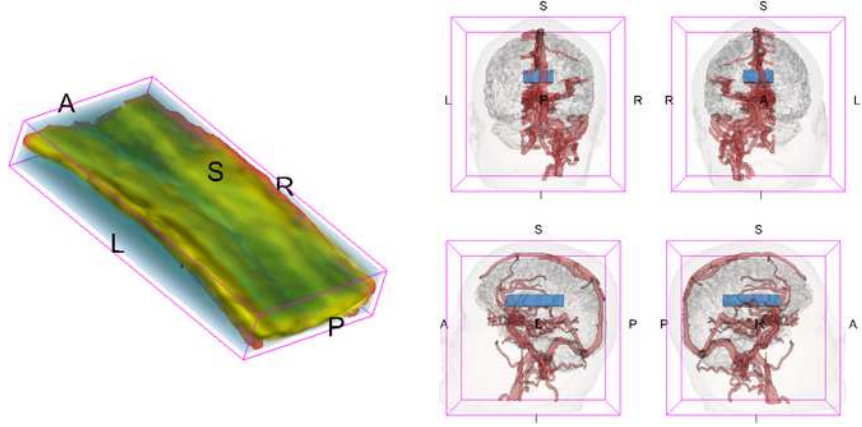}

\caption{{The 3D-rendering of the region of interest (left), a
blue block containing corpus callosum, and the template brain (right).}
Views: R${}={}$Right, L${}={}$Left, S${}={}$Superior, I${}={}$Interior, A${}={}$Anterior,
P${}={}$Posterior. For the purposes of orientation, major venous structures
are displayed in red in the right half of the template brain. The
3D-renderings are obtained using \citet{3D-Slicer:2011} and 3D
reconstructions of the anatomy from \citet{Pujol:2010}.}
\label{figure: 3D-rendering}
\end{figure}%
%f2 #&#
\begin{figure}[b]

\includegraphics{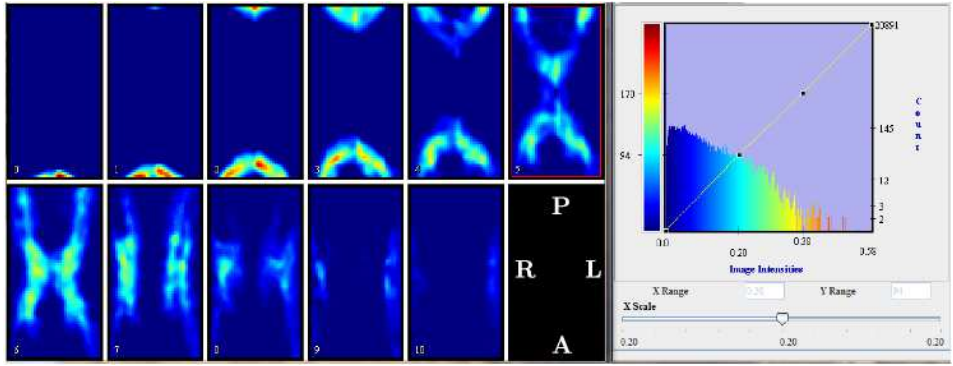}

\caption{The corpus callosum of a randomly chosen subject. Eleven axial
slices are shown on the left. A histogram of the weighted FA values is
on the right. Orientation: Interior (slice $0$) to Superior (slice $10$),
Posterior (top) to Anterior (bottom), Right to Left. The pictures are
obtained using \citet{MIPAV:2011}.}
\label{figure: subject_01_visit_01}
\end{figure}

The goal of the study was to quantify the location and size of
longitudinal variability of FA along the corpus callosum. The primary
region of interest (ROI) is a central block of the brain containing the
corpus callosum, the major bundle of neural fibers connecting the left
and right cerebral hemispheres. We weight FA at each voxel in the block
with a probability for the voxel to be in the corpus callosum, where
the probability is derived from an atlas formed using healthy-volunteer
scans, and study longitudinal changes of weighted FAs in the blocks
[\citet{Reich:etal:2010}]. Figure~\ref{figure: 3D-rendering} displays
the ROI {that contains corpus callosum together with its
relative location} in a template brain. Each block is of size $38\times
72 \times11$, indicating that there are $38$ sagittal, $72$ coronal,
and $11$ axial slices, respectively. Figure~\ref{figure:
subject_01_visit_01} displays the $11$ axial (horisontal) slices for
one of the subjects from bottom to top. In this paper, we study the FA
at every voxel of the blue blocks, which could be unfolded into an
approximately $30\mbox{,}000$ dimensional vector that contains the
corresponding FA value at each entry. The variability of these images
over multiple visits and subjects will be described by the combination
of the following: (1)~a~subject-specific imaging random intercept that
quantifies the cross-sectional variability; (2) a subject-specific
imaging slope that quantifies the dynamic irreversible deformation over
multiple visits; and (3) a subject-visit-specific imaging deviation that
quantifies exchangeable or reversible visit-to-visit changes.

%%
%0.35]{embeded_cc_22.png} \\
%0.05]{view_sc_1.png}
%%
%0.25]{view_sc_2.png}
%0.25]{view_sc_3.png}\\
%0.25]{view_sc_4.png}
%0.25]{view_sc_5.png}
%%
%

%0.65]{11_slices.png}
%axial slices are shown on the left. A histogram of the weighted FA
%values is on the right. Orientation: Interior(slice $0$) to
%Superior(slice $10$), Posterior (top) to Anterior(bottom), Right to
%Left. The pictures are obtained using \citet{MIPAV:2011}.}

%%
%0.65]{11_slices.png}
%%[width=10cm,grid,tics=10]
%%
%
%
% other approaches:(wavelets/smoothing group).
High-dimensional data sets have motivated the statistical and imaging
communities to develop new methodological approaches to data analysis.
Successful modeling approaches involving wavelets and splines and
adaptive kernels have been reported in the literature
[\citet{Bigelow:Dunson:2009},
\citet{Guo:2002},
\citet{Hua:etal:2012},
\citet{Li:etal:2011},
\citet{Mohamed:Davatzikos:2004},
\citet{Morris:and:Carrol:2006},
\citet{Morris:etal:2011},
\citeauthor{Reiss:Ogden:2008a} (\citeyear{Reiss:Ogden:2008a,Reiss:Ogden:2010}),
\citet{Reiss:etal:2005},
\citet{Rodriguez:etal:2009},
\citet{Yuan:etal:2014},
\citet{Zhu:etal:2011}].
A different direction of research has
focused on principal component decompositions [\citet
{Chong-Zhi:etal:2008,Crainiceanu:etal:2009,Aston:etal:2010,Staicu:etal:2010,Greven:etal:2010,Chong-Zhi:etal:2010,Zipunnikov:etal:2011a,Crainiceanu:etal:2010a}],
which led to several
applications to imaging data [\citet
{Shinohara:etal:2011,Goldsmith:etal:2011,Zipunnikov:etal:2011b}].
However, the high
dimensionality of new data sets, the inherent complexity of sampling
designs and data collection, and the diversity of new technological
measurements raise multiple challenges that are currently unaddressed.

% how we are going to deal with it: HD-LFPCA
{Here we address the problem of exploring and analyzing
populations of high-dimensional images at multiple visits using
high-dimensional longitudinal functional principal components analysis
(HD-LFPCA).}
The method decomposes the longitudinal imaging data into
subject-specific, longitudinal subject-specific, and
subject-visit-specific components. The dimension reduction for all
components is done using principal components of the corresponding
covariance operators. Note that we are interested in imaging
applications and do not perform smoothing. However, in Section~\ref
{subsec:HD-LFPCA-WN}, we discuss how the proposed approach can be
paired with smoothing and applied to high-dimensional functional data.
The estimation and inferential methods are fast and can be performed on
standard personal computers to analyze hundreds or thousands of
high-dimensional images at multiple visits. This was achieved by the
following combination of statistical and computational methods: (1)
relying only on matrix block calculations and sequential access to
memory to avoid loading very large data sets into the computer memory
[see \citet{Demmel:1997} and \citet{GolubVanLoan:1996} for a
comprehensive review of partitioned matrix techniques]; (2) using SVD
for matrices that have at least one dimension smaller than $10\mbox{,}000$
[\citet{Zipunnikov:etal:2011b}]; (3) obtaining best linear unbiased
predictors (BLUPs) of principal scores as a by-product of SVD of the
data matrix; and (4) linking the high-dimensional space to a
low-dimensional intrinsic space, which allows Karhunen--Lo\`{e}ve (KL)
decompositions of covariance operators that cannot even be stored in
the computer memory. Thus, the proposed methods are computationally
linear in the dimension of images.

The rest of the manuscript is organized as follows. Section~\ref
{section:LFPCA} reviews LFPCA and discusses its limitation in
high-dimensional settings. In Section~\ref{sec:HD-LFPCA} we introduce
HD-LFPCA, which provides a new statistical and computational framework
for LFPCA. This will circumvent the problems associated with LFPCA in
high-dimensional settings. Simulation studies are provided in
Section~\ref{sec:Simulations}. Our methods are applied to the MS data in
Section~\ref{sec:Application}. Section~\ref{sec:Discussion} concludes
the paper with a discussion.

%s2 #&#
\section{Longitudinal FPCA}\label{section:LFPCA}
In this section we review the LFPCA framework introduced by \citet
{Greven:etal:2010}. We develop an estimation procedure based on the
original one in \citet{Greven:etal:2010}, but we heavily modify it to
make it practical for applications to imaging high-dimensional data. We
also present the major reasons why the original methods cannot be
applied to high-dimensional data.

%s2.1 #&#
\subsection{Model}\label{subsection:LFPCA-model}

A brain imaging longitudinal study usually contains a sample of images
$\bY_{ij}$, where $\bY_{ij}$ is a recorded brain image of the $i$th
subject, $i=1,\ldots, I$, scanned at times $T_{ij}, j=1,\ldots,J_i$.
The total number of subjects is denoted by $I$. The times $T_{ij}$ are
subject specific. Different subjects could have a different number of
visits (scans), $J_i$. The images are stored in 3-dimensional array
structures of dimension $p=p_1\times p_2 \times p_3$. For example, in
the MS data $p = 38\times72 \times11 = 30\mbox{,}096$. Note that our
approach is not limited to the case when data are in a 3-dimensional
array. Instead, it can be applied directly to any data structure where
the voxels (or pixels, or locations, etc.) are the same across subjects
and visits, and data can be unfolded into a vector. {Following
\citet{Greven:etal:2010}, we consider the LFPCA model
%
%e1 #&#
\begin{equation}
\label{eq:lfpca-01} Y_{ij}(v) = \eta(v) + X_{i,0}(v) +
X_{i,1}(v)T_{ij} + W_{ij}(v),
\end{equation}
where $v$ denotes a voxel, $\eta(v)$ is a fixed main effect,
$X_{i,0}(v)$ is the random imaging intercept for subject $i$,
$X_{i,1}(v)$ is the random imaging slope for subject $i$, $T_{ij}$ is
the time of visit $j$ for subject $i$, $W_{ij}(v)$ is the random
subject/visit-specific imaging deviation. For simplicity, the main
effect $\eta(\cdot)$ does not depend on $i$ and $j$.} As discussed in
\citet{Greven:etal:2010}, model (\ref{eq:lfpca-01}) and the more general
model (\ref{eq:general-model-01}) in Section~\ref
{subsec:the-gen-fun-mix-model} are similar to functional models with
uncorrelated [\citet{Guo:2002}] and correlated [\citet
{Morris:and:Carrol:2006}] random functional effects. Instead of using
smoothing splines and wavelets as in \citet
{Guo:2002,Morris:and:Carrol:2006}, our approach models the covariance structures
using functional principal component analysis; we have found this
approach to lead to the major computational advantages, as further
discussed in Section~\ref{sec:HD-LFPCA}.

In the remainder of the paper, we unfold the data $\bY_{ij}$
and represent it as a $p\times1$ dimensional vector containing the
voxels in a particular order, where the order is preserved across all
subjects and visits. We assume that $\eta(v)$ is a fixed surface/image
and the latent (unobserved) bivariate process $X_i(v) =
(X^{\prime}_{i,0}(v), X^{\prime}_{i,1}(v))^{\prime}$ and process $W_{ij}(v)$ are
square-integrable stochastic processes. We also assume that $X_i(v)$
and $W_{ij}(v)$ are uncorrelated. We denote by $\bK^X(v_1,v_2)$ and
$\bK^W(v_1,v_2)$ their covariance operators, respectively. Assuming
that $\bK^X(v_1,v_2)$ and $\bK^W(v_1,v_2)$ are continuous, we can use
the standard Karhunen--Lo\`{e}ve expansions of the random processes
[\citet{Karhunen:1947,Loeve:1978}] and represent $X_{i}(v) = \sum_{k=1}^{\infty}\xi_{ik}\phi_{k}^{X}(v)$ with $\phi_{k}^{X}(v)=(\phi
_{k}^{X,0}(v),\phi_{k}^{X,1}(v))$ and $W_{ij}(v)=\sum_{l=1}^{\infty
}\zeta_{ijl}\phi_{l}^{W}(v)$,
where $\phi_{k}^{X}$ and $\phi_{l}^{W}$ are the eigenfunctions of the
$\bK^X$ and $\bK^W$ operators, respectively. Note that $\bK^X$ and
$\bK
^W$ will be estimated by their sample counterparts on finite $2p\times
2p$ and $p\times p$ grids, respectively. Hence, we can always make a
working assumption of continuity for $\bK^X$ and $\bK^W$. The LFPCA
model becomes the mixed effects model
%
%e2 #&#
\begin{equation}
\label{eq:model-04} \cases{ %
\displaystyle Y_{ij}(v)=\eta(v)+
\bZ^{\prime}_{ij}\sum_{k=1}^{\infty}
\xi_{ik}\phi _{k}^{X}(v)+\sum
_{l=1}^{\infty}\zeta_{ijl}\phi_{l}^{W}(v),
\vspace*{2pt}\cr
(\xi_{ik_1},\xi_{ik_2}) \sim\bigl(0,0;\lambda_{k_1}^{X},
\lambda _{k_2}^{X},0\bigr);\qquad (\zeta_{ijl_1},
\zeta_{ijl_2}) \sim\bigl(0,0;\lambda _{l_1}^{W},
\lambda_{l_2}^{W},0\bigr),}
\end{equation}
where $\bZ_{ij} = (1,T_{ij})^{\prime}$ and ``$\sim(0,0;\lambda
_{k_1}^{X},\lambda_{k_2}^{X},0)$'' indicates that a pair of variables is
uncorrelated with mean zero and variances $\lambda_{k_1}^{X}$ and
$\lambda_{k_2}^{X}$, respectively. Variances $\lambda^{X}_k$'s are
nonincreasing, that is, $\lambda^X_{k_1}\ge\lambda^X_{k_2}$ if $k_1
\le k_2$. {We do not require normality of the scores in the
model. The only assumption is the existence of second order moments of
the distribution of scores.} In addition, the assumption that $X_i(v)$
and $W_{ij}(v)$ are uncorrelated is ensured by the assumption that $\{
\xi_{ik}\}_{k=1}^{\infty}$ and $\{\zeta_{ijl}\}_{l=1}^{\infty}$ are
uncorrelated. Note that model (\ref{eq:model-04}) may be extended to
include a more general vector of covariates $\bZ_{ij}$. We discuss a
general functional mixed model in Section~\ref{subsec:the-gen-fun-mix-model}.

In practice, model \ref{eq:model-04} is projected onto the first $N_X$
and $N_W$ components of $\bK^X $ and $\bK^W$, respectively. Assuming
that $N_X$ and $N_W$ are known, the model becomes
%
%e3 #&#
\begin{equation}
\label{eq:model-05} \cases{ %
\displaystyle  Y_{ij}(v)=\eta(v)+
\bZ^{\prime}_{ij}\sum_{k=1}^{N_X}
\xi_{ik}\phi _{k}^{X}(v)+\sum
_{l=1}^{N_W}\zeta_{ijl}\phi_{l}^{W}(v),
\vspace*{2pt}\cr
(\xi_{ik_1},\xi_{ik_2}) \sim\bigl(0,0;\lambda_{k_1}^{X},
\lambda _{k_2}^{X},0\bigr);\qquad (\zeta_{ijl_1},
\zeta_{ijl_2}) \sim\bigl(0,0;\lambda _{l_1}^{W},
\lambda_{l_2}^{W},0\bigr).}
\end{equation}
The choice of the number of principal components $N_X$ and $N_W$ is
discussed in \citet{Chong-Zhi:etal:2008,Greven:etal:2010}. Typically,
$N_X$ and $N_W$ are small and (\ref{eq:model-05}) provides significant
dimension reduction of the family of images and their longitudinal
dynamics. The main reason why the LFPCA model (\ref{eq:model-05})
cannot be fit when data are high dimensional is that the empirical
covariance matrices $\bK^X$ and $\bK^W$ cannot be calculated, stored,
or diagonalized. Indeed, in our case these operators would be $30\mbox{,}000$
by $30\mbox{,}000$ dimensional, which would have around $1$ billion entries.
In other applications these operators would be even bigger.

%s2.2 #&#
\subsection{Estimation}\label{subsection:LFPCA-estimation}
{Our estimation is based on the methods of moments (MoM) for
pairwise quadratics $E(\bY_{ij_1}\bY_{kj_2}^{\prime})$. The computationally
intensive part of fitting~(\ref{eq:model-05}) is estimating the
following massively multivariate model:
%
%e4 #&#
\begin{eqnarray}
\label{eq:01} \bY_{ij} &=& \eta+ \sum_{k=1}^{N_X}
\xi_{ik}{\bolds\phi}^{X,0}_k+T_{ij}
\sum_{k=1}^{N_X}\xi_{ik}{\bolds\phi}^{X,1}_k +\sum
_{l=1}^{N_W}\zeta_{ijl}{\bolds\phi}^W_l
\nonumber
\\[-8pt]
\\[-8pt]
\nonumber
& = &\eta+\bPhi^{X,0}
\bxi_i+T_{ij}\bPhi^{X,1}\bxi_i +
\bPhi^W\bzeta_{ij},
\end{eqnarray}
where $\eta= (\eta(v_1), \ldots, \eta(v_p))$, $\bY_{ij} = \{
Y_{ij}(v_1),\ldots, Y_{ij}(v_p)\}$ are $p\times1$ dimensional vectors,
${\bolds\phi}^{X,0}_k$, ${\bolds\phi
}^{X,1}_k$, and ${\bolds\phi}^W_l$ are correspondingly
vectorized eigenvectors, $\bPhi^{X,0}=[{\bolds\phi
}^{X,0}_1,\ldots,{\bolds\phi}
^{X,0}_{N_X}]$ and $\bPhi^{X,1}=[{\bolds\phi
}^{X,1}_1,\ldots,{\bolds\phi}
^{X,1}_{N_X}]$ are $p\times N_X$ dimensional matrices, $\bPhi
^{W}=[{\bolds\phi}
^{W}_1,\ldots,{\bolds\phi}^{W}_{N_W}]$ is a $p\times
N_W$ dimensional matrix,
principal scores $\bxi_{i}=(\xi_{i1},\ldots,\xi_{iN_X})^{\prime}$ and
$\bzeta
_{ij}=(\zeta_{ij1},\ldots, \zeta_{ijN_U})^{\prime}$ are uncorrelated with
diagonal covariance matrices $E(\bxi_i\bxi_i^{\prime}) = \bLambda^X= {\rm
diag}(\lambda^X_1, \ldots, \lambda^X_{N_X})$ and $E(\bzeta
_{ij}\bzeta
_{ij}^{\prime}) = \bLambda^W = {\rm diag}(\lambda^W_1, \ldots, \lambda
^W_{N_W})$, respectively.

To obtain the eigenvectors and eigenvalues in model (\ref{eq:01}), the
spectral decompositions of $\bK^X$ and $\bK^W$ need to be constructed.
The first $N_X$ and $N_W$ eigenvectors and eigenvalues are retained
after this, that is, $\bK^X \approx\bPhi^X\bLambda^X\bPhi^{X'}$ and
$\bK^W \approx\bPhi^W\bLambda^W\bPhi^{W'}$, where $\bPhi^X =
[\bPhi
^{X,0'}, \bPhi^{X,1'}]^{\prime}$ denotes a $2p \times N_X$ matrix with
orthonormal columns and $\bPhi^W$ is a $p\times N_W$ matrix with
orthonormal columns.

\begin{lemma}\label{le1} The MoM estimators of the covariance operators and
the mean in~(\ref{eq:01}) are unbiased and given by
%
%e5 #&#
\begin{eqnarray}
\label{eq:OLS-03} %\begin{array}{l}
\hat{\bK}_X^{00}& = &\sum
_{i,j_1,j_2}\bbY_{ij_1}\bbY _{ij_2}^{\prime}h^{1}_{ij_1j_2},\qquad
\hat{\bK}_X^{01} = \sum_{i,j_1,j_2}
\bbY_{ij_1}\bbY _{ij_2}^{\prime}h^{2}_{ij_1j_2},\nonumber\\
\hat{\bK}_X^{10} &=& \sum_{i,j_1,j_2}
\bbY_{ij_1}\bbY _{ij_2}^{\prime}h^{3}_{ij_1j_2},
\qquad
\hat{\bK}_X^{11} = \sum_{i,j_1,j_2}
\bbY_{ij_1}\bbY _{ij_2}^{\prime}h^{4}_{ij_1j_2},\\
 \hat{\bK}^W &=& \sum_{i,j_1,j_2}
\bbY_{ij_1}\bbY _{ij_2}^{\prime}h^{5}_{ij_1j_2},\qquad
\hbeta= \frac{1}{n}\sum_{i=1}^I\sum
_{j=1}^{J_i}\bY_{ij},\nonumber
\end{eqnarray}
where $\bbY_{ij} = \bY_{ij}-\hbeta$, the $2p\times2p$ matrix $\bK
^X =
[\bK_X^{00}\vdots\bK_X^{01};\bK_X^{10}\vdots\bK_X^{11}]$, with
$\bK
_X^{ks}=E\{\bPhi^{X,k}\bxi_i(\bPhi^{X,s}\bxi_i)^{\prime}\}$ for $k,s \in
\{
0,1\}$, the weights $h^l_{ij_1j_2}$ are elements of the $l${th} column
of the matrix $\bH_{m\times5} = \bF^{\prime}(\bF\bF^{\prime})^{-1}$, the matrix
$\bF_{5\times m}$ has columns equal to $\bff
_{ij_1j_2}=(1,T_{ij_2},T_{ij_1},T_{ij_1}T_{ij_2},\delta_{j_1j_2})^{\prime}$,
and $m = \sum_{i=1}^IJ_i^2$.
\end{lemma}

The proof of the lemma is given in the \hyperref[app]{Appendix}. The MoM estimators
(\ref{eq:OLS-03}) define the symmetric matrices $\hat{\bK}^X$ and
$\hat
{\bK}^W$. Identifiability of model (\ref{eq:01}) requires that some
subjects have more than two visits, that is, $J_i\ge3$. Note that if
one is only interested in estimating covariances, $\eta$ can be
eliminated as a nuisance parameter by using MoMs for quadratics of
differences $E(\bY_{ij_1}-\bY_{kj_2})(\bY_{ij_1}-\bY_{kj_2})^{\prime}$
as in
\citet{Shou:etal:2013}.

Estimating the covariance matrices is a crucial first step. However,
constructing and storing these matrices requires $O(p^2)$ calculations
and $O(p^2)$ memory units. Even if it were possible to calculate and
store these covariances, obtaining the spectral decompositions would be
infeasible. Indeed, $\bK^X$ is a $2p\times2p$ and $\bK^W$ is a
$p\times p$ dimensional matrix, which would require $O(p^3)$
operations, making diagonalization infeasible for $p > 10^4$.
Therefore, LFPCA, which performs well when the functional
dimensionality is moderate, fails in very high and
ultrahigh-dimensional settings.

In the next section we develop a methodology capable of handling
longitudinal models of very high dimensionality. The main reason why
these methods work efficiently is because the intrinsic dimensionality
of the model is controlled by the sample size of the study, which is
much smaller compared to the number of voxels. The core part of the
methodology is to carefully exploit this underlying low-dimensional space.}

%s3 #&#
\section{HD-LFPCA}\label{sec:HD-LFPCA}
In this section we provide our statistical model and inferential
methods. The main emphasis is on providing a new methodological
approach with the ultimate goal of solving the intractable
computational problems discussed in the previous section.

%s3.1 #&#
\subsection{Eigenanalysis}\label{subsec:HD-LFPCA-Eigenanalysis}
\label{subsec:eigenanalysis}

In Section~\ref{section:LFPCA} we established that the main
computational bottleneck for standard LFPCA of \citet{Greven:etal:2010}
is constructing, storing, and decomposing the relevant covariance
operators. In this section we propose an algorithm that allows
efficient calculation of the eigenvectors and eigenvalues of these
covariance operators {without either calculating or storing the
covariance operators.} In addition, we demonstrate how all necessary
calculations can be done using sequential access to data. One of the
main assumptions of this section is that the sample size, $n=\sum_{j=1}^IJ_i$,
is moderate, so calculations of order $O(n^3)$ are
feasible. In Section~\ref{sec:Discussion} we discuss ways to
extend our approach to situations when this assumption is violated.

Write $\bbY=(\bbY_1,\ldots,\bbY_I)$, where $\bbY_i = (\bbY
_{i1},\ldots,\bbY_{iJ_i})$ is a centered $p\times J_i$ matrix and the column $j$,
$j=1,\ldots,J_i$, contains the unfolded image for subject $i$ at visit~$j$.
Note that the matrix $\bbY_i$ contains all the data for subject
$i$ with each column corresponding to a particular visit. The matrix
$\bbY$ is the $p\times n$ matrix obtained by
column-binding\vspace*{1pt} the centered subject-specific data matrices $\bbY_i$.
Thus, if $\bbY_{i}=(\bbY_{i1},\ldots,\bbY_{iJ_i})$, then $\bbY
=(\bbY
_1,\ldots,\bbY_I)$. Our approach starts with constructing the SVD of
the matrix $\bbY$:
%
%e6 #&#
\begin{equation}
\label{eq:eigenanalysis-02-b} \bbY= \bV\bS^{1/2}\bU^{\prime}.
\end{equation}
Here, the matrix $\bV$ is $p\times n$ dimensional with $n$ orthonormal
columns, $\bS$ is a diagonal $n\times n$ dimensional matrix, and $\bU$
is an $n\times n$ dimensional orthogonal matrix.
Calculating the SVD of $\bbY$ requires only a number of operations
linear in the number of parameters $p$. Indeed, consider the $n\times
n$ symmetric matrix $\bbY^{\prime}\bbY$ with its spectral decomposition\vspace*{1pt}
$\bbY
^{\prime}\bbY=\bU\bS\bU^{\prime}$. Note that for high-dimensional $p$ the matrix
$\bbY$ cannot be loaded into the memory. The solution is to partition
it into $L$ slices as $\bbY^{\prime} = [(\bbY^1)^{\prime}| (\bbY^2)^{\prime}|
\cdots
|(\bbY^L)^{\prime}]$, where the size of the $l$th slice, $\bbY^l$, is $(p/L)
\times n$ and can be adapted to the available computer memory and
optimized to reduce implementation time. The matrix $\bbY^{\prime}\bbY$ is
then calculated as $\sum_{l=1}^L(\bbY^l)^{\prime}\bbY^l$ by streaming the
individual blocks. This step calculates singular value decomposition of
the $p\times n$ matrix $\tilde{\mathbf{Y}}$. Note that for any permutation
of components $v$, model (\ref{eq:model-05}) will be valid and the
covariance structure imposed by the model can be recovered by doing the
inverse permutation. If smoothing of the covariance matrix is
desirable, then this step can be efficiently combined with Fast
Covariance Estimation [FACE, \citet{Xiao:etal:2013}], a computationally
efficient smoother of (low-rank) high-dimensional covariance matrices
with $p$ up to 100,000.

From the SVD (\ref{eq:eigenanalysis-02-b}) the $p\times n$ matrix $\bV$
can be obtained as $\bV=\bbY\bU\bS^{-1/2}$. The actual calculations can
be performed on the slices of the partitioned matrix $\bbY$ as $\bV
^l=\bbY^l\bU\bS^{-1/2}, l=1,\ldots,L$. The concatenated slices
$[(\bV
^1)^{\prime}|\break  (\bV^2)^{\prime}| \cdots|(\bV^L)^{\prime}]$ form the matrix of the left
singular vectors $\bV^{\prime}$. Therefore, the SVD (\ref
{eq:eigenanalysis-02-b}) can be constructed with sequential access to
the data $\bbY$ with $p$-linear effort.

After obtaining the SVD of $\bbY$, each image $\bbY_{ij}$ can be
represented as
$\bbY_{ij} = \bV\bS^{1/2}\bU_{ij}$, where $\bU_{ij}$ is a corresponding
column of matrix $\bU^{\prime}$. Therefore, the vectors $\bbY_{ij}$ differ
only through the vector factors $\bU_{ij}$ of dimension $n\times1$.
Comparing this SVD representation of $\bbY_{ij}$ with the right-hand
side of (\ref{eq:01}), it follows that \emph{cross-sectional and
longitudinal variability controlled by the principal scores $\bxi_i$,
$\bzeta_{ij}$, and time variables $T_{ij}$ must be completely
determined by the low-dimensional vectors $\bU_{ij}$}. This is the key
observation which makes the approach feasible. Below, we provide more
intuition behind our approach. The formal argument is presented in
Lemma~\ref{le2}.

First, we substitute the left-hand side of (\ref{eq:01}) with its SVD
representation of $\bbY_{ij}$ to get $\bV\bS^{1/2}\bU_{ij} = \bPhi
^{X,0}\bxi_i+T_{ij}\bPhi^{X,1}\bxi_i +\bPhi^W\bzeta_{ij}$.
Now we can multiply by $\bV'$ both sides of the equation to get
$\bS^{1/2}\bU_{ij} = \bV'\bPhi^{X,0} \bxi_i + T_{ij}\bV'\bPhi
^{X,1} \bxi
_i + \bV'\bPhi^W \bzeta_{ij}$.
If we denote $\bA^{X,0} = \bV'\bPhi^{X,0}$ of size $n\times N_X$,
$\bA
^{X,1} = \bV'\bPhi^{X,1}$ of size $n\times N_X$, and $\bA^W = \bV
'\bPhi
^U$ of size $n\times N_W$, we obtain
%
%e7 #&#
\begin{equation}
\label{eq:04} \bS^{1/2}\bU_{ij} = \bA^{X,0}
\bxi_i + T_{ij}\bA^{X,1} \bxi_i +
\bA^W \bzeta_{ij}.
\end{equation}
Conditionally on the observed data, $\bbY$, models (\ref{eq:01}) and
(\ref{eq:04}) are equivalent. Indeed, model (\ref{eq:01}) is a \emph
{linear} model for the $n$ vectors $\bbY_{ij}$'s. These vectors span an
(at most) $n$-dimensional linear subspace. Hence, the columns of the
matrix $\bV$, the right singular vectors of $\bbY$, could be thought of
as an orthonormal basis, while $\bS^{1/2}\bU_{ij}$ are the coordinates
of $\bbY_{ij}$ in this basis. Multiplication by $\bV'$ can be seen as a
linear mapping from model (\ref{eq:01}) for the high-dimensional
observed data $\bbY_{ij}'s$ to model~(\ref{eq:04}) for the
low-dimensional data $\bS^{1/2}\bU_{ij}$. Additionally, even though
$\bV
\bV' \neq\bI_p$, the projection defined by $\bV$ is lossless in the
sense that model (\ref{eq:01}) can be recovered from model (\ref
{eq:04}) using the identity $\bV\bV'\bbY_{ij}=\bbY_{ij}$. Hence, model
(\ref{eq:04}) has an ``intrinsic'' dimensionality induced by the study
sample size, $n$. We can estimate the low-dimensional model (\ref
{eq:04}) using the LFPCA methods described in Section~\ref
{section:LFPCA}. This step is now feasible, as it requires only
$O(n^3)$ calculations. The formal result presented below shows that
fitting model (\ref{eq:04}) is an essential step for getting the
high-dimensional principal components in $p$-linear time.

 \begin{lemma}\label{le2} The eigenvectors of the estimated covariance
operators (\ref{eq:OLS-03}) can be calculated as
$ \hat{\bPhi}^{X,0} = \bV\bhA^{X,0}, \hat{\bPhi}^{X,1} =
\bV\bhA
^{X,1}, \hat{\bPhi}^W = \bV\bhA^W$,
where the matrices $\bhA^{X,0}$, $\bhA^{X,1}$, $\bhA^W$ are obtained
from fitting model (\ref{eq:04}). The estimated matrices of eigenvalues
$\hat{\bLambda}^X$ and $\hat{\bLambda}^W$ are the same for both model
(\ref{eq:01}) and model (\ref{eq:04}).
\end{lemma}

The proof of the lemma is given in the \hyperref[app]{Appendix}. This result is a
generalization of the HD-MFPCA result in \citet{Zipunnikov:etal:2011a},
which was obtained in the case when there is no longitudinal component
$\bPhi^{X,1}$. In the next section we provide more insights into the
intrinsic model (\ref{eq:04}).

%s3.2 #&#
\subsection{The general functional mixed model}\label
{subsec:the-gen-fun-mix-model}
A natural way to generalize model~(\ref{eq:model-05}) is to
consider the following model:
%
%e8 #&#
\begin{eqnarray}
\label{eq:general-model-01} \bY_{ij} &=& \eta+ Z_{ij,0}\sum
_{k=1}^{N_X}\xi_{ik}{\bolds\phi}
^{X,0}_k+Z_{ij,1}\sum_{k=1}^{N_X}
\xi_{ik}{\bolds\phi }^{X,1}_k+\cdots
\nonumber
\\[-8pt]
\\[-8pt]
\nonumber
&&{}
+Z_{ij,q}\sum_{k=1}^{N_X}
\xi_{ik}{\bolds\phi}^{X,q}_k +\sum
_{l=1}^{N_W}\zeta _{ijl}{\bolds\phi}^W_l,
\end{eqnarray}
where the $(q+1)$-dimensional vector of covariates $\bZ
_{ij}=(Z_{ij,0},Z_{ij,1},\ldots,Z_{ij,q})$ may include, for instance,
polynomial terms of $T_{ij}$ and other covariates of interest.

The fitting approach is essentially the same as the one described for
the LFPCA model in Section~\ref{subsec:eigenanalysis}. As before, the
right singular vectors $\bU_{ij}$ contain the longitudinal information
about $\bxi_i$, $\bzeta_i$, and covariates $\bZ_{ij}$. The following
two results are direct generalizations of Lemmas \ref{le1} and \ref{le2}.

\begin{lemma}\label{le3}The MoM estimators of the covariance operators and
the mean in~(\ref{eq:general-model-01}) are unbiased and given by
\begin{eqnarray*}
\hat{\bK}_X^{ks} &=& \sum
_{i,j_1,j_2}\bbY_{ij_1}\bbY _{ij_2}^{\prime}h^{1+s+k(q+1)}_{ij_1j_2},\qquad
 \hat{\bK}^W = \sum_{i,j_1,j_2}
\bbY_{ij_1}\bbY _{ij_2}^{\prime}h^{(q+1)^2+1}_{ij_1j_2},
\\
\hbeta&=& \frac{1}{n}\sum_{i=1}^I\sum
_{j=1}^{J_i}\bY_{ij},
\end{eqnarray*}
where $\bbY_{ij} = \bY_{ij}-\hbeta$, the $(q+1)p\times(q+1)p$
block-matrix $\bK^X$ is composed of $p\times p$ matrices $\bK
_X^{ks}=E\{
\bPhi^{X,k}\bxi_i(\bPhi^{X,s}\bxi_i)^{\prime}\}$ for $k,s \in\{
0,1,\ldots
,q\}$, the weights $h^l_{ij_1j_2}$ are elements of the $l${th} column
of matrix $\bH_{m\times((q+1)^2+1)} = \bF^{\prime}(\bF\bF^{\prime})^{-1}$, the
matrix $\bF_{((q+1)^2+1)\times m}$ has columns equal to $\bff
_{ij_1j_2}=(\operatorname{vec}(\bZ_{ij_1}\otimes\bZ_{ij_2}),\delta_{j_1j_2})^{\prime}$, and
$m = \sum_{i=1}^IJ_i^2$.
\end{lemma}

\begin{lemma}\label{le4} The eigenvectors of the estimated covariance
operators for (\ref{eq:general-model-01}) can be calculated as
$ \hat{\bPhi}^{X,k} = \bV\bhA^{X,k}, k = 0,1,\ldots,q, \hat
{\bPhi
}^W = \bV\bhA^W$,
where the matrices $\bhA^{X,k}, k = 0,1,\ldots,q $, and $\bhA^W$ are
obtained from fitting the intrinsic model
%
%e9 #&#
\begin{eqnarray}
\label{eq:general-model-02} \bS^{1/2}\bU_{ij} &=& Z_{ij,0}\sum
_{k=1}^{N_X}\xi_{ik}\bA
^{X,0}_k+Z_{ij,1}\sum_{k=1}^{N_X}
\xi_{ik}\bA^{X,1}_k+\cdots
\nonumber
\\[-8pt]
\\[-8pt]
\nonumber
&&{}+Z_{ij,q}\sum
_{k=1}^{N_X}\xi_{ik}
\bA^{X,q}_k +\sum_{l=1}^{N_W}
\zeta _{ijl}\bA^W_l.
\end{eqnarray}
The estimated matrices of eigenvalues $\hat{\bLambda}^X$ and $\hat
{\bLambda}^W$ are the same for both model~(\ref{eq:general-model-01})
and model (\ref{eq:general-model-02}).
\end{lemma}

%s3.3 #&#
\subsection{Estimation of principal scores}\label{subsec:est-of-prin-scores}

{The principal scores are the coordinates of $\bbY_{ij}$ in the
basis defined by the LFPCA model (\ref{eq:general-model-01}).} In this
section we propose an approach to calculating BLUP of the scores that
is computationally feasible for samples of high-resolution images.

First, we introduce some notation. In Section~\ref
{subsec:HD-LFPCA-Eigenanalysis} we showed that the SVD of the matrix
$\bbY$ can be written as $\bbY_{i}=\bV\bS^{1/2}\bU_{i}^{\prime}$, where the
$n\times J_i$ matrix $\bU_{i}^{\prime}$ corresponds to the subject $i$.
Model (\ref{eq:general-model-01}) can be rewritten as
%
%e10 #&#
\begin{equation}
\label{eq:blups-ols} \operatorname{vec}(\bbY_i) = \bB_i
\bomega_i,
\end{equation}
where $\bB_i = [\bB_i^X \vdots\bB_i^W]$, $\bB_i^X = \bZ
_{i,0}\otimes
\bPhi^{X,0}+\bZ_{i,1}\otimes\bPhi^{X,1} + \cdots+ \bZ
_{i,q}\otimes
\bPhi^{X,q}$, $\bB_i^W = \bI_{J_i}\otimes\bPhi^{W}$, $\bZ
_{i,k}=(Z_{i1,k},\ldots,Z_{iJ_i,k})^{\prime}$, $\bomega_i=(\bxi
_i^{\prime},\bzeta
_i^{\prime})^{\prime}$, the subject level principal scores
$\bzeta_i=(\bzeta^{\prime}_{i1},\ldots,\bzeta_{iJ_i}^{\prime})^{\prime}$, $\otimes
$ is
the Kronecker product of matrices, and operation $\operatorname{vec}(\cdot)$ stacks
the columns of a matrix on top of each other.
The following lemma contains the main result of this section; it shows
how the estimated BLUPs can be calculated for the LFPCA model.

\begin{lemma}\label{le5} Under the general LFPCA model (\ref
{eq:general-model-01}), the estimated best linear unbiased predictor
(EBLUP) of $\bxi_i$ and $\bzeta_i$ is given by
%
%e11 #&#
\begin{equation}
\label{eq:blups-01} %
\pmatrix{ \hat{\bxi}_{i}\vspace*{2pt}
\cr
\hat{\bzeta}_{i} } %
= \bigl(\hat{\bB}_i^{\prime}
\hat{\bB}_i\bigr)^{-1}\hat{\bB}_i^{\prime}\operatorname{vec}(
\bbY_i),
\end{equation}
where all matrix factors on the right-hand side can be written in terms of
the low-dimensional right singular vectors.
\end{lemma}

The proof of the lemma is given in the \hyperref[app]{Appendix}. The EBLUPs
calculations are almost instantaneous, as the matrices involved in
(\ref
{eq:blups-01}) are low-dimensional and do not depend on the dimension
$p$. Section~\ref{subsec:large-sample-size} in the \hyperref[app]{Appendix} briefly
describes how the framework can be adapted to settings with tens or
hundreds of thousands images.

%s3.4 #&#
\subsection{HF-LFPCA model with white noise}\label{subsec:HD-LFPCA-WN}
{The original LFPCA model in \citet{Greven:etal:2010} was
developed for functional observations and contained an additional white
noise term. In this section, we show how the HD-LFPCA framework can be
extended to accommodate such a term and how the extended model can be estimated.

We now seek to fit the following model:
%
%e12 #&#
\begin{eqnarray}
\label{eq:general-model-03} \bY_{ij} &= &\eta+ Z_{ij,0}\sum
_{k=1}^{N_X}\xi_{ik}{\bolds\phi}
^{X,0}_k+Z_{ij,1}\sum_{k=1}^{N_X}
\xi_{ik}{\bolds\phi }^{X,1}_k+\cdots
\nonumber
\\[-8pt]
\\[-8pt]
\nonumber
&&{}
+Z_{ij,q}\sum_{k=1}^{N_X}
\xi_{ik}{\bolds\phi}^{X,q}_k +\sum
_{l=1}^{N_W}\zeta _{ijl}{\bolds\phi}^W_l + \varepsilon_{ij},
\end{eqnarray}
where $\varepsilon_{ij}$ is a $p$-dimensional white noise variable, that
is, $E(\varepsilon_{ij}) = 0_p$ for any $i,j$ and $E(\varepsilon
_{i_1j_1}\varepsilon_{i_2j_2})=\sigma^2\delta_{i_1i_2}\delta
_{j_1j_2}\bI
_p$. The white noise process $\varepsilon_{ij}(v)$ is assumed to be
uncorrelated with processes $X_i(v)$ and $W_{ij}(v)$.

Lemma~\ref{le3} applied to (\ref{eq:general-model-03}) shows that $\hat
{K}^W_{\sigma^2} =\sum_{i,j_1,j_2}\bbY_{ij_1}\bbY
_{ij_2}^{\prime}h^{(q+1)^2+1}_{ij_1j_2}$ is an unbiased estimator of $K^W +
\sigma^2\bI_p$. To estimate $\sigma^2$ in a functional
case, we can follow the method in \citet{Greven:etal:2010}: (i) drop
the diagonal elements of $\hat{K}^W_{\sigma^2}$ and use a bivariate
smoother to get $\tilde{K}^W_{\sigma^2}$, (ii) calculate an estimator
$\hat{\sigma}^2 = \max\{(\operatorname{tr}(\hat{K}^W_{\sigma^2})-\operatorname{tr}(\tilde
{K}^W_{\sigma
^2})/p,0\}$.
To make this approach feasible in very high-dimensional settings
($p\sim100\mbox{,}000$), we can use the fast covariance estimation (FACE)
developed in \citet{Xiao:etal:2013}, a bivariate smoother that scales
up linearly with respect to $p$ and preserves the low dimensionality of
the estimated covariance operator. Thus, HD-LFPCA remains feasible
after smoothing by FACE.

When the observations $\bY_{ij}$'s are nonfunctional, the off-diagonal
smoothing approach cannot be used. In this case, if one assumes that
model (\ref{eq:general-model-03}) is low-rank, then $\sigma^2$ can be
estimated as $(\operatorname{tr}(\hat{K}^W_{\sigma^2})-\sum_{k=1}^{N_W}\hat
{\lambda
}_k^W)/(p-N_W)$. Bayesian model selection approaches that estimate both
the rank of PCA models and variance $\sigma^2$ are discussed in \citet
{Everson:Roberts:2000} and \citet{Minka:2000}.
}

%s4 #&#
\section{Simulations}\label{sec:Simulations}
In this section three simulation studies are used to explore the
properties of our proposed methods. In the first study, we replicate
several simulation scenarios in \citet{Greven:etal:2010} for functional
curves, but we focus on using a number of parameters up to two orders
of magnitude larger than the ones in the original scenarios. This
increase in dimensionality could not be handled by the original LFPCA
approach. In the second study, we explore how methods recover $3$D
spatial bases when the approach of \citet{Greven:etal:2010} cannot be
implemented. In the third study, we replicate the unbalanced design in
and use time variable $T_{ij}$ from our DTI application and generate
data using principal components estimated in Section~\ref
{sec:Application}. For each scenario, we simulated $100$ data sets. All
three studies were run on a four core i7-2.67~GHz PC with 6~Gb of RAM
memory using Matlab 2010a. The software is available upon request.

\textit{First scenario \textup{(1D,} functional curves\textup{)}}. We follow \citet
{Greven:etal:2010} and generate data as follows:
\[
\cases{ %
\displaystyle Y_{ij}(v)=\sum
_{k=1}^{N_X}\xi_{ik}\phi_{k}^{X,0}(v)+T_{ij}
\sum_{k=1}^{N_X}\xi_{ik}
\phi_{k}^{X,1}(v)+\sum_{l=1}^{N_W}
\zeta _{ijl}\phi _{l}^{W}(v)+\varepsilon_{ij}(v),\vspace*{2pt}\cr
\qquad v\in\mcV,
\vspace*{2pt}\cr
\xi_{ik} \stackrel{\mathrm{i.i.d.}} {\sim} 0.5N\Bigl(-\sqrt{
\lambda _{k}^{X}}/2,\lambda _{k}^{X}/2
\Bigr)+0.5N\Bigl(\sqrt{\lambda_{k}^{X}}/2,
\lambda_{k}^{X}/2\Bigr),
\vspace*{2pt}\cr
\zeta_{ijl} \stackrel{\mathrm{i.i.d.}} {\sim} 0.5N\Bigl(-\sqrt
{\lambda _{l}^{W}}/2,\lambda_{l}^{W}/2
\Bigr)+0.5N\Bigl(\sqrt{\lambda_{l}^{W}}/2,\lambda
_{l}^{W}/2\Bigr), }
\]
where $\xi_{ik} \stackrel{\mathrm{i.i.d.}}{\sim} 0.5N(-\sqrt{\lambda
_{k}^{X}}/2,\lambda_{k}^{X}/2)+0.5N(\sqrt{\lambda_{k}^{X}}/2,\lambda
_{k}^{X}/2)$ means that the scores $\xi_{ik}$ are simulated from a
mixture of two normals, $N(-\sqrt{\lambda_{k}^{X}}/2,\lambda
_{k}^{X}/2)$ and $N(\sqrt{\lambda_{k}^{X}}/2,\lambda_{k}^{X}/2)$ with
equal probabilities; a similar notation holds for $\zeta_{ijl}$. The
scores $\xi_{ik}$'s and $\zeta_{ijl}$'s are mutually independent. We
set $I=100$, $J_i=4, i=1,\ldots,I$, and the number of eigenfunctions
$N_X=N_W=4$. The true eigenvalues are the same, $\lambda_k^{X}=\lambda
_k^{W}=0.5^{k-1}, k = 1,2,3,4$. The orthogonal but not mutually
orthogonal bases were
\begin{eqnarray*}
\phi^{X,0}_1(v) &=&
\sqrt{2/3}\sin(2\pi v), \qquad \phi^{X,1}_1(v) = 1/2, \qquad
\phi^{W}_1 = \sqrt{4}\phi^{X,1}_1,
\\
\phi^{X,0}_2(v)& =&\sqrt{2/3}\cos(2\pi v), \qquad
\phi^{X,1}_2(v) = \sqrt {3}(2v-1)/2, \qquad \phi^{W}_2
= \sqrt{4/3}\phi^{X,0}_1,
\\
\phi^{X,0}_3(v) &=& \sqrt{2/3}\sin(4\pi v), \qquad
\phi^{X,1}_3(v) = \sqrt {5}\bigl(6v^2-6v+1
\bigr)/2, \\
 \phi^{W}_3 &=& \sqrt{4/3}\phi^{X,0}_2,
\\
\phi^{X,0}_4(v) &=& \sqrt{2/3}\cos(4\pi v), \qquad
\phi^{X,1}_4(v) = \sqrt {7}\bigl(20v^3-30v^2+12v-1
\bigr)/2, \\
 \phi^{W}_4 &=& \sqrt{4/3}\phi^{X,0}_3,
\end{eqnarray*}
which are measured on a regular grid of $p$ equidistant points in the
interval $[0,1]$. To explore scalability, we consider several grids
with an increasing number of sampling points, $p$, equal to $ 750,
3000$, $12\mbox{,}000, 24\mbox{,}000, 48\mbox{,}000$, and $96\mbox{,}000$. Note that a brute-force
extension of the standard LFPCA would be at the edge of feasibility for
such a large $p$. For each $i$, the first time $T_{i1}$ is generated
from the uniform distribution over interval $(0,1)$ denoted by $U(0,1)$.
Then differences $(T_{ij+1}-T_{ij})$ are also generated from $U(0,1)$
for $1\le j\le3$. The times $T_{i1},\ldots,T_{i4}$ are normalized to
have sample mean zero and variance one. Although no measurement noise
is assumed in model (\ref{eq:model-05}), we simulate data that also
contains white noise, $\varepsilon_{ij}(v)$. The purpose of this is
twofold. First, it is of interest to explore how the presence of white
noise affects the performance of methods which do not model it
explicitly. Second, the choice of the eigenfunctions in the original
simulation scenario of \citet{Greven:etal:2010} makes the estimation
problem ill-posed if data does not contain white noise. The white noise
$\varepsilon_{ij}(v)$ is assumed to be i.i.d. $N(0,\sigma^2)$ for each
$i,j,v$ and independent of all other latent processes. To evaluate
different signal-to-noise ratios, we consider values of $\sigma^2$
equal to $0.0001, 0.0005, 0.001, 0.005, 0.01$. Note that we normalized
each of the data generating eigenvectors to have norm one. Thus, the
signal-to-noise ratio, $(\sum_{k=1}^{4}\lambda_k^{X}+\sum_{k=1}^{4}\lambda_k^{W})/(p\sigma^2)$, ranges from $50$ (for $p=750$
and $\sigma^2 = 0.0001$) to $0.004$ (for $p=96\mbox{,}000$ and $\sigma^2 = 0.01$).

Table~\ref{table:sim2-phi-x0} and Tables~1 and~2 in the online supplement [\citet
{Zipunnikov:etal:supplement:2014}] report the average $L_2$ distances
between the estimated and true eigenvectors for $X_{i,0}(v)$,
$X_{i,1}(v)$, and $W_{ij}(v)$, respectively. The averages are
calculated based on $100$ simulated data sets for each $(p, \sigma^2) $
combination. Standard deviations are shown in brackets. Three trends
are obvious: (i) eigenvectors with larger eigenvalues are estimated
with higher accuracy, (ii) larger white noise corresponds to a
decreasing accuracy, (iii) for identical levels of white noise,
accuracy goes down when the dimension $p$ goes up. Similar trends are
observed for average distances between estimated and true eigenvalues
reported in Tables~3 and~4. These trends follow from the fact that for any
fixed $\sigma^2$, the signal-to-noise ratio decreases with increasing
$p$ and the performance of the approach quickly deteriorates once the
signal-to-noise ratio becomes smaller than one.

%t1 #&#
\begin{table}
\tabcolsep=0pt
\caption{Based on 100 simulated data sets, average distances between
estimated and true eigenvectors of $X_{i,0}(v)$; standard deviations
are given in parentheses}
\label{table:sim2-phi-x0}
\begin{tabular*}{\textwidth}{@{\extracolsep{\fill}}ld{1.10}d{1.10}d{1.10}d{1.10}@{}}
\hline
$\bolds{(p$,  $\sigma^2)}$ & \multicolumn{1}{c}{$\bolds{\|\phi^{X,0}_1 - \hat
{\phi
}^{X,0}_1\|^2}$} & \multicolumn{1}{c}{$\bolds{\|\phi^{X,0}_2 - \hat{\phi
}^{X,0}_2\|^2}$} & \multicolumn{1}{c}{$\bolds{\|\phi^{X,0}_3 - \hat{\phi
}^{X,0}_3\|^2}$} & \multicolumn{1}{c@{}}{$\bolds{\|\phi^{X,0}_4 - \hat{\phi
}^{X,0}_4\|^2}$} \\
\hline
(750, 1e--04) &0.034\ (0.048) &0.07\ (0.069) &0.074\ (0.053) &0.081\ (0.07) \\
(750, 5e--04) &0.031\ (0.031) &0.055\ (0.051) &0.084\ (0.097) &0.112\ (0.151) \\
(750, 0.001) &0.035\ (0.039) &0.062\ (0.054) &0.078\ (0.059) &0.139\ (0.206) \\
(750, 0.005) &0.035\ (0.039) &0.072\ (0.062) &0.096\ (0.063) &0.159\ (0.084) \\
(750, 0.01) &0.045\ (0.036) &0.079\ (0.054) &0.129\ (0.102) &0.234\ (0.103) \\[3pt]
(3000, 1e--04) &0.031\ (0.028) &0.064\ (0.118) &0.09\ (0.13) &0.109\ (0.126) \\
(3000, 5e--04) &0.037\ (0.032) &0.065\ (0.048) &0.077\ (0.06) &0.14\ (0.136) \\
(3000, 0.001) &0.031\ (0.027) &0.06\ (0.044) &0.087\ (0.062) &0.131\ (0.07) \\
(3000, 0.005) &0.058\ (0.035) &0.106\ (0.058) &0.171\ (0.09) &0.324\ (0.096) \\
(3000, 0.01) &0.073\ (0.028) &0.142\ (0.048) &0.236\ (0.074) &0.508\ (0.072) \\[3pt]
(12\mbox{,}000, 1e--04) &0.031\ (0.028) &0.062\ (0.048) &0.077\ (0.056) &0.134\ (0.165) \\
(12\mbox{,}000, 5e--04) &0.041\ (0.036) &0.078\ (0.05) &0.121\ (0.069) &0.201\ (0.081) \\
(12\mbox{,}000, 0.001) &0.047\ (0.04) &0.083\ (0.054) &0.164\ (0.114) &0.295\ (0.118) \\
(12\mbox{,}000, 0.005) &0.112\ (0.032) &0.217\ (0.064) &0.44\ (0.216) &0.758\ (0.153) \\
(12\mbox{,}000, 0.01) &0.175\ (0.031) &0.338\ (0.093) &0.554\ (0.132) &0.987\ (0.071) \\[3pt]
(24\mbox{,}000, 1e--04) &0.035\ (0.032) &0.066\ (0.049) &0.09\ (0.141) &0.146\ (0.173) \\
(24\mbox{,}000, 5e--04) &0.055\ (0.045) &0.097\ (0.061) &0.146\ (0.09) &0.266\ (0.098) \\
(24\mbox{,}000, 0.001) &0.07\ (0.038) &0.125\ (0.047) &0.23\ (0.167) &0.43\ (0.15) \\
(24\mbox{,}000, 0.005) &0.183\ (0.049) &0.348\ (0.097) &0.622\ (0.208) &0.998\ (0.11) \\
(24\mbox{,}000, 0.01) &0.295\ (0.043) &0.518\ (0.117) &0.742\ (0.102) &1.184\ (0.07) \\[3pt]
(48\mbox{,}000, 1e--04) &0.046\ (0.068) &0.076\ (0.067) &0.103\ (0.059) &0.175\ (0.122) \\
(48\mbox{,}000, 5e--04) &0.073\ (0.035) &0.13\ (0.056) &0.234\ (0.1) &0.437\ (0.099) \\
(48\mbox{,}000, 0.001) &0.105\ (0.051) &0.183\ (0.065) &0.407\ (0.23) &0.695\ (0.192) \\
(48\mbox{,}000, 0.005) &0.307\ (0.08) &0.532\ (0.151) &0.824\ (0.208) &1.19\ (0.086) \\
(48\mbox{,}000, 0.01) &0.458\ (0.084) &0.712\ (0.1) &0.938\ (0.074) &1.186\ (0.126) \\[3pt]
(96\mbox{,}000, 1e--04) &0.045\ (0.033) &0.087\ (0.059) &0.146\ (0.103) &0.246\ (0.107) \\
(96\mbox{,}000, 5e--04) &0.116\ (0.081) &0.194\ (0.094) &0.431\ (0.268) &0.721\ (0.218) \\
(96\mbox{,}000, 0.001) &0.188\ (0.089) &0.32\ (0.121) &0.787\ (0.339) &1.062\ (0.216) \\
(96\mbox{,}000, 0.005) &0.457\ (0.065) &0.707\ (0.107) &0.954\ (0.125) &1.298\ (0.074) \\
(96\mbox{,}000, 0.01) &0.662\ (0.105) &0.926\ (0.103) &1.116\ (0.075) &1.143\ (0.153) \\
\hline
\end{tabular*}
\end{table}

\begin{figure}

\includegraphics{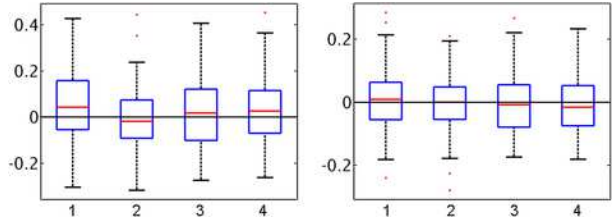}

\caption{Boxplots of the normalized estimated eigenvalues for process
$X_{i}(v)$, $(\hat{\lambda}^{X}_{k}-\lambda^{X}_k)/\lambda^{X}_k$ (left
box), and the normalized estimated eigenvalues for process $W_{ij}(v)$,
$(\hat{\lambda}^{W}_{l}-\lambda^{W}_l)/\lambda^{W}_l$ (right box),
based on scenario $1$ with $100$ replications. The zero is shown by the
solid black line.}
\label{figure:scenario-01-lambda}
\end{figure}

\begin{figure}[b]

\includegraphics{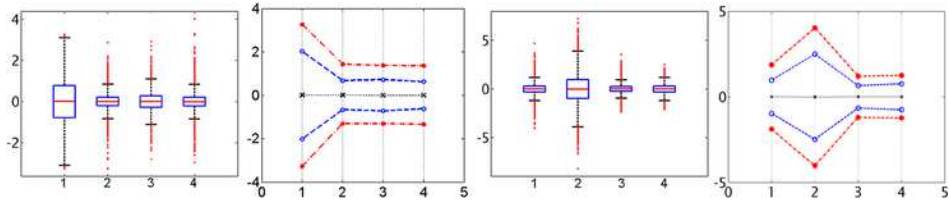}

\caption{The left two panels show the distribution of the normalized
estimated scores of process $X_{i}(v)$, $(\xi_{ik}-\hat{\xi
}_{ik})/\sqrt
{\lambda^{X}_k}$. Boxplots are given in the left column. The right
column shows the medians (black marker), $5\%$ and $95\%$ quantiles
(blue markers), and $0.5\%$ and $99.5\%$ quantiles (red markers).
Similarly, the distribution of the normalized estimated scores of
process $W_{ij}(v)$, $(\zeta_{ijl}-\hat{\zeta}_{ijl})/\sqrt{\lambda
^{X}_l}$ is provided at the right two panels.}
\label{figure:scenario-01-diff}
\end{figure}

Figure 1 in the online supplement [Zipunnikov et al. (\citeyear{Zipunnikov:etal:supplement:2014})] displays the true and estimated eigenfunctions
for the case when $p = 12\mbox{,}000$ and $\sigma^2=0.01^2$ and shows the
complete agreement with Figure~2 in \citet{Greven:etal:2010}. The
boxplots of the estimated eigenvalues are displayed in Figure~\ref
{figure:scenario-01-lambda}. In Figure~\ref{figure:scenario-01-diff},
panels one and three report the boxplots of and panels two and four
display the medians and quantiles of the distribution of the normalized
estimated scores, $(\xi_{ik}-\hat{\xi}_{ik})/\sqrt{\lambda^{X}_{k}}$
and $(\zeta_{ijl}-\hat{\zeta}_{ijl})/\sqrt{\lambda^{W}_{l}}$,
respectively. This indicates that the estimation procedures provides
unbiased estimates.

%f4 #&#

%%
%%
%

%f5 #&#

\begin{figure}

\includegraphics{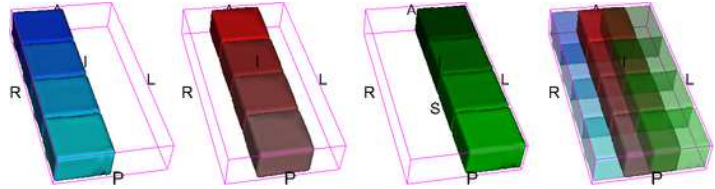}

\caption{3D eigenimages of the $2nd$ simulation scenario. From left to
right: $\phi^{X,0}_k$ are in blue, $\phi^{X,1}_k$ are in red, $\phi
^{W}_k$ are in green, the most right one shows the overlap of all
eigenimages. Views: R${}={}$Right, L${}={}$Left, S${}={}$Superior, I${}={}$Interior,
A${}={}$Anterior, P${}={}$Posterior. The 3D-renderings are obtained using \citet
{3D-Slicer:2011}.}
\label{figure: scenario-02-phi-3D-renderings}
\end{figure}

\textit{Second scenario} (3D). Data sets in this study replicate the 3D
ROI blocks from the DTI MS data set. We simulated $100$ data sets from
the model
\[
\cases{ %
\displaystyle Y_{ij}(v)=\sum
_{k=1}^{N_X}\xi_{ik}\phi_{k}^{X,0}(v)+T_{ij}
\sum_{k=1}^{N_X}\xi_{ik}
\phi_{k}^{X,1}(v)+\sum_{l=1}^{N_W}
\zeta _{ijl}\phi _{l}^{W}(v),\qquad v\in\mcV,
\vspace*{2pt}\cr
\xi_{ik} \stackrel{\mathrm{i.i.d.}} {\sim} N\bigl(0,
\lambda_{k}^{X}\bigr)\quad \mbox{and}\quad \zeta_{ijl}
\stackrel{\mathrm{i.i.d.}} {\sim} N\bigl(0,\lambda_{l}^{W}
\bigr), }
\]
where $\mcV= [1,38]\times[1,72]\times[1,11]$. Eigenimages ($\phi
^{X,0}_k$, $\phi^{X,1}_k$) and $\phi^W_{l}$ are displayed in Figure~\ref
{figure: scenario-02-phi-3D-renderings}. The images in this scenario
can be thought of as 3D images with voxel intensities on the $[0,1]$
scale. The voxels within each sub-block (eigenimage) are set to $1$ and
outside voxels are set to $0$. There are four blue and red sub-blocks
corresponding to $\phi^{X,0}_k$ and $\phi^{X,1}_k$, respectively. The
eigenfunctions closest to the anterior side of the brain (labeled A in
Figure~\ref{figure: 3D-rendering}) are $\phi^{X,0}_1$ and $\phi^{X,1}_1$, which have the
strongest signal proportional to the largest eigenvalue (variance),
$\lambda^X_1$. The eigenvectors that are progressively closer to the
posterior part of the brain (labeled~P) correspond to smaller
eigenvalues represented as lighter shades of blue and red,
respectively. The sub-blocks closest to the P have the smallest signal,
which is proportional to $\lambda^X_4$. The eigenimages $\phi^W_k$
shown in green are ordered the same way. Note that $\phi^{X,0}_k$ are
uncorrelated with $\phi^{W}_l$. However, both $\phi^{X,0}_k$ and
$\phi
^{W}_l$ are correlated with the $\phi^{X,1}_k$'s describing the random
slope $X_{i,1}(v)$. We assume that $I=150$, $J_i=6, i=1,\ldots,I$, and
the true eigenvalues $\lambda_k^{X}=0.5^{k-1}, k = 1,2,3$, and
$\lambda
_l^{W}=0.5^{l-1}, l = 1,2$. The times $T_{ij}$ were generated as in the
first simulation scenario. To apply HD-LFPCA, we unfold each image $\bY
_{ij}$ and obtain vectors of size $p = 38\times72\times11 = 30\mbox{,}096$.
The entire simulation study took $20$ minutes or approximately $12$
seconds per data set.
%
%f3 #&#
%
%0.36]{scenario-02-phi_x0.png}
%0.36]{scenario-02-phi_x1.png}
%0.36]{scenario-02-phi_u.png}
%0.36]{scenario-02-phi_all.png}
%
%%
Figures~4, 5 and~6 in the
online supplement [\citet{Zipunnikov:etal:supplement:2014}] display the
medians of the estimated eigenimages and the voxelwise $5$th and $95$th
percentile images, respectively. All axial slices, or $z$ slices in a
$x$--$y$--$z$ coordinate system, are the same. Therefore, we display only one
$z$-slice, which is representative of the entire 3D image. To obtain a
grayscale image with voxel values in the $[0,1]$ interval, each
estimated eigenvector, $\hat{{\bolds\phi}}= ({\hat
{\phi}_1},\ldots, {\hat{\phi
}_p})$, was normalized as $\hat{{\bolds\phi}}
\rightarrow(\hat{{\bolds\phi}} - \min_s{\hat{\phi}_s})/(\max_s{\hat{\phi}_s}-\min_s{\hat{\phi}_s})$.
Figure 4 in the online supplement [Zipunnikov et al. (\citeyear{Zipunnikov:etal:supplement:2014})] displays the voxelwise
medians of
the estimator, indicating that the method recovers the spatial
configuration of both bases. The $5$-percentile and $95$-percentile
images are displayed in Figures 5 and 6 in the online supplement [Zipunnikov et al. (\citeyear{Zipunnikov:etal:supplement:2014})], respectively. Overall, the original
pattern is recovered with some small distortions most likely due to the
correlation between bases (please note the light gray patches).

The boxplots of the estimated normalized eigenvalues $(\hat{\lambda
}^{X}_{k}-\lambda^{X}_k)/\lambda^{X}_k$ and $(\hat{\lambda
}^{W}_{l}-\lambda^{W}_l)/\lambda^{W}_l$ are displayed in
Figure 2 in the online supplement [Zipunnikov et al. (\citeyear{Zipunnikov:etal:supplement:2014})].
The eigenvalues are estimated
consistently. However, in $6$ out of $100$ cases (extreme values shown
in red), the estimation procedure did not distinguish well between
$\phi
^{W}_3$ and $\phi^{W}_4$. This is probably due the relatively low signal.

The boxplots of the estimated eigenscores are displayed in
Figure 3 in the online supplement [Zipunnikov et al. (\citeyear{Zipunnikov:etal:supplement:2014})].
In this scenario, the total number of the
estimated scores $\xi_{ik}$ is $15\mbox{,}000$ for each $k$ and there are
$90\mbox{,}000$ estimated scores $\zeta_{ijl}$ for each $l$.
The distributions of the normalized estimated scores $(\xi_{ik}-\hat
{\xi
}_{ik})/\sqrt{\lambda^{X}_{k}}$ and $(\zeta_{ijl}-\hat{\zeta
}_{ijl})/\sqrt{\lambda^{W}_{l}}$ are displayed in the first and third
panels of Figure 3 in the online supplement
[Zipunnikov et al. (\citeyear{Zipunnikov:etal:supplement:2014})], respectively. The
spread of the distributions increases as the signal-to-noise ratio
decreases. The second and fourth panels of Figure 3 in the online supplement
[Zipunnikov et al. (\citeyear{Zipunnikov:etal:supplement:2014})] display the medians, $0.5\%$, $5\%$, $95\%$,
and $99.5\%$ quantiles of the distribution of the normalized estimated scores.

\textit{Third scenario \textup{(3D}, empirical basis\textup{)}}. We generate data
using the first ten principal components estimated in Section~\ref
{sec:Application}.
We replicated the unbalanced design of the MS study and used the same
time variable $T_{ij}$'s. The principal scores $\xi_{ik}$ and $\zeta
_{ijk}$ were simulated as in Scenario 1 with $\lambda_k^{X}=\lambda
_k^{W}=0.5^{k-1}, k = 1,\ldots, 10$. The white noise variance $\sigma
^2$ was set to $10^{-4}$. Thus, SNR is equal to 1.32. The results are
reported in Table~5 in the online supplement [\citet
{Zipunnikov:etal:supplement:2014}].
The average distances between estimated and true eigenvectors for
$X_{i}(v)$ and $W_{ij}(v)$ are calculated based on $100$ simulated data
sets. Principal components and principal scores become less accurate as
the signal-to-noise gets smaller.

%%
%
%%
%

%s5 #&#
\section{Longitudinal analysis of brain fractional anisotropy in MS
patients}\label{sec:Application}

{In this section we apply HD-LFPCA to the DTI images of MS
patients. The study population included individuals with no, mild,
moderate, and severe disability. Over the follow-up period (as long as
5 years in some cases), there was little change in the median
disability level of the cohort. Cohort characteristics are reported in
Table~7 in the online supplement
[\citet{Zipunnikov:etal:supplement:2014}]. The scans have been aligned
using a $12$ degrees of freedom transformation, meaning that we
accounted for rotation, translation, scaling, and shearing, but not for
nonlinear deformation. As described in Section~\ref{sec:Introduction},
the primary region of interest is a central block of the brain of size
$38 \times72 \times11$ displayed in Figure~\ref{figure: 3D-rendering}.
We weighted each voxel in the block with a probability for the voxel to
be in the corpus callosum and study longitudinal changes of weighted
voxels in the blocks [\citet{Reich:etal:2010}]. Probabilities less than
$0.05$ were set to zero. Below we model longitudinal variability of the
weighted FA at every voxel of the blocks. The entire analysis performed
in Matlab 2010a took only $3$ seconds on a PC with a quad core
i7-2.67~GHz processor and 6~Gb of RAM memory. First, we unfolded each
block into a $30\mbox{,}096$ dimensional vector that contained the
corresponding weighted FA values. In addition to high dimensionality,
another difficulty of analyzing this study was the unbalanced
distribution of scans across subjects (see Table~6 in the online supplement [\citet
{Zipunnikov:etal:supplement:2014}]); this is a typical problem in
natural history studies. After forming the data matrix $\bY$, we
estimated the overall mean as $\hat{\eta}=\frac{1}{n}\sum_{i=1}^I\sum_{j=1}^{J_i}\bY_{ij}$ and de-meaned the data. {The estimated
mean is shown in Figure 7 in the online supplement [Zipunnikov et al. (\citeyear{Zipunnikov:etal:supplement:2014})].
The mean
image across subjects and visits indicates a shape characterized by our
scientific collaborators as a ``standard corpus callosum template.''}

\textit{Model} 1: First, we start by fitting a random intercept and random
slope model~(\ref{eq:lfpca-01}). To enable comparison of the
variability explained by processes $X_{i}(v)$ and $W_{ij}(v)$, we
followed the normalization procedure in Section~3.4 in \citet
{Greven:etal:2010}: $T_{ij}$'s were normalized to have sample mean zero
and sample variance one.
The estimated covariance matrices are not necessarily nonnegative
definite. Indeed, we have obtained small negative eigenvalues of the
covariance operators $\bhK^X$ and $\bhK^W$. Following \citet
{Hall:etal:2008}, all the negative eigenvalues were set to zero. The
total variation was decomposed into the ``subject-specific'' part
modeled by process $X_i$ and the ``exchangeable visit-to-visit'' part
modeled by the process $W_{ij}$. Most of the total variability, $70.8\%
$, is explained by $X_i$ (subject-specific variability) with the trace
of $\bK^X = 122.53$, while $29.2\%$ is explained by $W_{ij}$
(exchangeable visit-to-visit variability) with the trace of $\bK^W =
50.47$. Two major contributions of our approach are to separate the
processes $X_i$ and $W_{ij}$ and quantify their corresponding
contributions to the total variability.
%
%t2 #&#
\begin{table}
\caption{Model 1 ($T_{ij}$ change): Cumulative variability explained by
the first $10$ eigenimages}\label{table:data-analysis-cumulative-var-explained}
\begin{tabular*}{\textwidth}{@{\extracolsep{\fill}}ld{2.2}ccc@{}}
\hline
\multicolumn{1}{@{}l}{$\bolds{k}$} & \multicolumn{1}{c}{$\bolds{\phi_k^{X,0}}$} & $\bolds{\phi_k^{X,1}}$ & $\bolds{\phi_k^{W}}$ & \multicolumn{1}{c@{}}{\textbf{Cumulative}} \\
\hline
\phantom{0}1 & 22.13 & 0.08 & 7.12 & 29.33 \\
\phantom{0}2 & 10.66 & 0.11 & 3.20 & 43.29 \\
\phantom{0}3 & 5.99 & 0.13 & 2.04 & 51.44 \\
\phantom{0}4 & 4.84 & 0.08 & 1.44 & 57.80 \\
\phantom{0}5 & 2.80 & 0.06 & 0.90 & 61.56 \\
\phantom{0}6 & 2.39 & 0.07 & 0.83 & 64.85 \\
\phantom{0}7 & 1.94 & 0.10 & 0.63 & 67.52 \\
\phantom{0}8 & 1.72 & 0.08 & 0.50 & 69.82 \\
\phantom{0}9 & 1.55 & 0.05 & 0.45 & 71.86 \\
10 & 1.20 & 0.05 & 0.39 & 73.50 \\[3pt]
& 55.20 & 0.80 & 17.50 & 73.50 \\
\hline
\end{tabular*}
\end{table}

%f6 #&#
\begin{figure}

\includegraphics{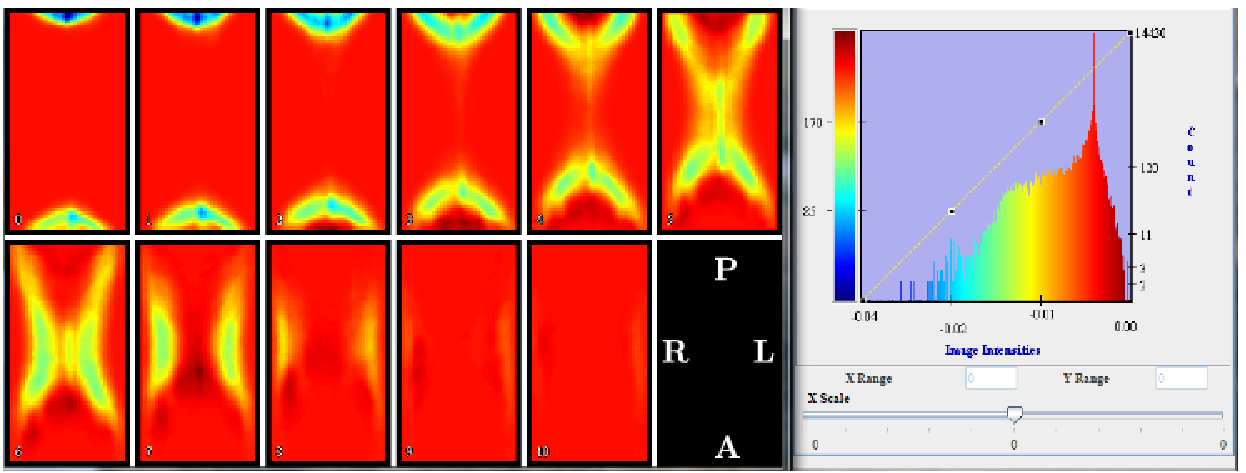}

\caption{Eleven slices of $\hat{\phi}^{X,0}_1$. A histogram of the
voxel intensities is on the right. The pictures are obtained using
\citet{MIPAV:2011}.}
\label{figure:data-analysis-phi-x0-1}
\end{figure}

%%
%0.65]{results-phi-x0-1.png}
%%[width=10cm,grid,tics=10]
%%
%
%

%f7 #&#
\begin{figure}[b]

\includegraphics{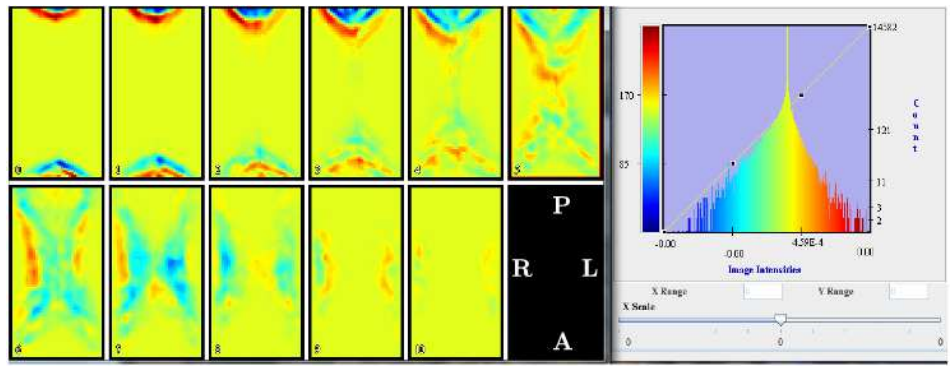}

\caption{Eleven slices of $\hat{\phi}^{X,1}_1$. A histogram of the
voxel intensities is on the right. The pictures are obtained using
\citet{MIPAV:2011}.}
\label{figure:data-analysis-phi-x1-1}
\end{figure}

Table~\ref{table:data-analysis-cumulative-var-explained} reports the
percentage explained by the first $10$ eigenimages. The first $10$
random intercept eigenimages explain roughly $55\%$ of the total
variability, while the effect of the random slope is accounting for
only $0.80\%$ of the total variability. The exchangeable variability
captured by $W_{ij}(v)$ accounts for $17.5\%$ of the total variation.

The first three estimated random intercept and slope eigenimages are
shown in pairs in Figures~\ref{figure:data-analysis-phi-x0-1}, \ref
{figure:data-analysis-phi-x1-1}, and in 8, 9,
10,
11 in the online supplement [\citet
{Zipunnikov:etal:supplement:2014}], respectively. Figures~12, 13 and
14 in the online supplement [\citet
{Zipunnikov:etal:supplement:2014}] display the first three eigenimages
of the exchangeable measurement error process $W_{ij}(v)$.
%The orientation of the slices follow the one in Figure~\ref{figure:
%subject_01_visit_01}.
%%
Each eigenimage is accompanied with the histogram of its voxel values.
Recall that the eigenimages were obtained by folding the unit length
eigenvectors of $p \approx3\cdot10^4$ voxels. Therefore, each voxel
is represented by a small value. For principal scores, negative and
positive voxel values correspond to opposite loadings (directions) of
variation. Each histogram has a peak at zero due to the existence of
the threshold for the probability maps indicating if a voxel is in the
corpus callosum. This peak is a convenient visual divider of the color
spectrum into the loading specific colors. Because of the sign
invariance of the SVD, the separation between positive and negative
loadings is comparable only within the same eigenimage. However, the
loadings of the random intercept and slope within an eigenimage of the
process $X_{i}(v)$ can be compared as they share the same principal
score. This allows us to contrast the time invariant random intercept
with the longitudinal random slope and, thus, to localize regions that
exhibit the largest longitudinal variability.
%This can potentially result in two patterns. In the first one, we
%would observe the overlapping regions of random slope and random
%intercept that share the same sign. In the second one, we would
%identify the overlapping regions which have the opposite signs.
%In brain imaging applications, these regions would get lighter or
%dimmer with time depending on the sign of the random slope and the
%principal score.
This could be used to analyze the longitudinal changes of brain
imaging in a particular disease or to help generate new scientific hypotheses.
%

%%
%0.65]{results-phi-x1-1.png}
%%[width=10cm,grid,tics=10]
%%
%

%0.65]{results-phi-x0-1.png}
%voxel intensities is on the right. The pictures are obtained using

%0.65]{results-phi-x1-1.png}
%voxel intensities is on the right. The pictures are obtained using
We now interpret the random intercept and slope parts of the
eigenimages obtained for the MS data. Figures~\ref
{figure:data-analysis-phi-x0-1} and\vspace*{1pt} \ref{figure:data-analysis-phi-x1-1}
show the random intercept and slope parts of the first eigenimage $\phi
^{X}_1$, respectively. The negatively loaded voxels of the random
intercept, $\phi^{X,0}_{1}$, essentially compose the entire corpus
callosum. This indicates an overall shift in the mean FA of the corpus
callosum. This is expected and is a widely observed empirical feature
of principal components. The random slope part, $\phi^{X,1}_{1}$, has
both positively and negatively loaded areas in the corpus callosum. The
areas colored in blue shades share the sign of the random intercept
$\phi^{X,0}_1$, whereas the red shades have the opposite sign. The
extreme colors of the spectrum of $\phi^{X,1}_{1}$ show a clear
separation into negative and positive loadings, especially accentuated
in the splenium (posterior) and the genu (anterior) areas of the corpus
callosum; please note the upper and lower areas in panels $0$ through
$5$ of Figure~\ref{figure:data-analysis-phi-x1-1}. This implies that a
subject with a positive first component score $\xi_{i1} > 0$ would tend
to have a smaller mean FA over the entire corpus callosum and the FA
would tend to decrease with time in the negatively loaded parts of the
splenium. The reverse will be true for a subject with a negative score
$\xi_{i1}$. The other two eigenimages of $X_i(v)$ and eigenimages of
$W_{ij}(v)$ are discussed in the online supplement [\citet
{Zipunnikov:etal:supplement:2014}].

Next, we explored whether the deviation process $W_{ij}(v)$ depends on
MS severity by analyzing the corresponding eigenscores. To do this, we
divided subjects according to their MS type into three subgroups:
relapsing-remitting (RR, 102 subjects), secondary progressive (SP, 40
subjects), and primary progressive (PP, 25 subjects). For each of the
first ten eigenimages, we formally tested whether there are differences
between the distributions of the scores of the three groups using the
$t$-test and the Mann--Whitney--Wilcoxon-rank test for equality of means
and the Kolmogorov--Smirnov test for equality of distributions. For the
first eigenimage, the scores in the SP group have been significantly
different from both those in RR and PP groups ($p$-values $<$0.005 for
all three tests). For the second eigenimage, scores in the RR group
were significantly different from both SP and PP ($p$-values $<$0.01
for all three tests). The two left images of Figure~\ref
{figure:model_1_W} display the group beanplots of the scores for the
first eigenimage and the second eigenimage of $W_{ij}(v)$, respectively.

%f8 #&#
\begin{figure}

\includegraphics{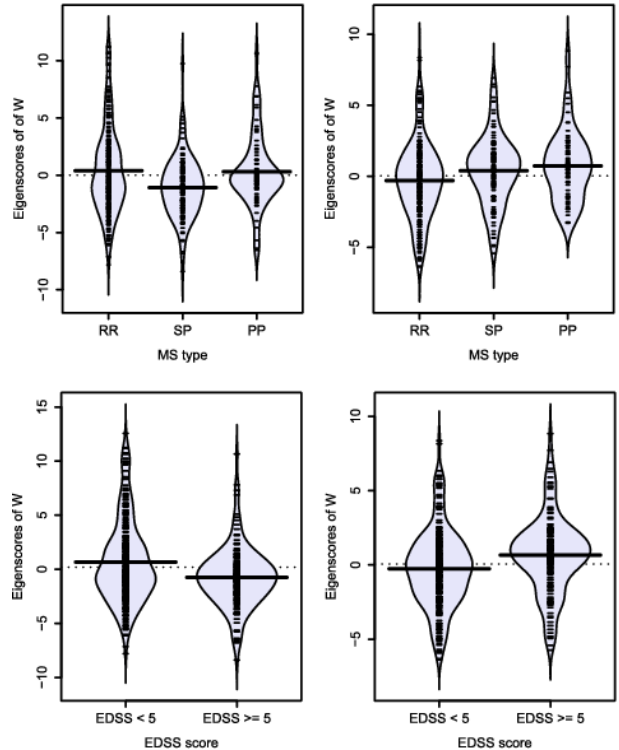}

\caption{Model 1: Group beanplots according to MS type (top) and
according to EDSS score (bottom).}
\label{figure:model_1_W}
\end{figure}

In addition to MS type, the EDSS scores were recorded at each visit. We
divided subjects into two groups according to their EDSS score: (i)
smaller than~5 and (ii)~larger than or equal to 5. As with MS type, we
have conducted tests for the equality of distributions of the
eigenscores of these two groups for all ten eigenimages. For
eigenimages one and two, the distributions of eigenscores have been
found to be significantly different ($p$-values $<$0.001 for all three
tests). The two right images on Figure~\ref{figure:model_1_W} display
group beanplots of the scores for the first eigenimage and the second
eigenimage of $W_{ij}(v)$, respectively.

We have also conducted a standard analysis based on the scalar mean FA
over the CC for each subject/visit and fitted a scalar random
intercept/random slope model. In this model, the random intercept
explains roughly $94\%$ of the total variation of the mean FAs.
Figure 15 in the online supplement [Zipunnikov et al. (\citeyear{Zipunnikov:etal:supplement:2014})]
displays beanplots of the estimated random
intercepts stratified by EDDS score and MS type. For both cases there
was a statistically significant difference between the distributions of
the random intercepts (EDSS: $p$-values $<0.001$; MS-type, SP vs. RR and
PP, $p$-values $< 0.002$, for all three tests). Similar tests for the
distributions of the random slopes did not identify statistically
significant differences between groups. We conclude that this simple
model agrees with the full HD-LFPCA mode, though the multivariate model
provides a detailed decomposition of the total FA variation together
with localization variability in the original 3D-space.

\textit{Model} 2. Second, we fit model (\ref{eq:general-model-01}) using
$Z_{ij,1}$ equal to a visit-specific EDSS score. Again, $Z_{ij,1}$'s
were normalized to have sample mean 0 and sample variance~1. Table~\ref
{table:data-analysis-model-2-cumulative-var-explained} reports
percentages explained by the first $10$ eigenimages in model~2.
Interestingly, the total variation explained by the random intercept
and random slope in both models is approximately the same, with 56.0\%
in model 1 vs. 54.2\% for model 2. However, the random slope in model 2
explains a much higher proportion of the total variation: 13.2\% in
model 2 using EDSS versus model 1 using time. The second component of
the random slope explains almost $8.5\%$ of the total variation. We
have also explored whether the scores of $W_{ij}(v)$ depend on MS type
and EDSS score using the $t$-test, the Mann--Whitney--Wilcoxon-rank test,
and the Kolmogorov--Smirnov test. For the first eigenimage, the SP type
was significantly different from the RR ($p$-values $<0.01$ for all
three tests), though it was not significantly different from the PP
group. For the second eigenimage, the distribution of eigenscores for
the SP type was significantly different from that of the scores for the
RR ($p$-values $<0.05$ for all three tests), and not significantly
different from the distribution of the scores of the PP type. For
grouping according to EDSS score, the distributions of the eigenscores
of the first two eigenimages have been found to be statistically
different ($p$-values $<0.01$ for all three tests). Figure~\ref
{figure:model_2_W} displays beanplots similar to Figure~\ref
{figure:model_1_W} for the distributions of the scores in the groups
defined by MS types and EDSS. This indicates that the deviation process
$W_{ij}(v)$ in models 1 and 2 carries not only useful but also almost
identical remaining information regarding severity of MS.

\begin{table}
\caption{Model 2 ($Z_{ij}$ change): Cumulative variability explained by
the first $10$ eigenimages}
\label{table:data-analysis-model-2-cumulative-var-explained}
\begin{tabular*}{\textwidth}{@{\extracolsep{\fill}}ld{2.2}ccc@{}}
\hline
\multicolumn{1}{@{}l}{$\bolds{k}$} & \multicolumn{1}{c}{$\bolds{\phi_k^{X,0}}$} & $\bolds{\phi_k^{X,1}}$ & $\bolds{\phi_k^{W}}$ & \multicolumn{1}{c@{}}{\textbf{Cumulative}} \\
\hline
\phantom{0}1 & 17.79 & 0.42 & 5.59 & 23.80 \\
\phantom{0}2 & 0.53 & 8.46 & 1.99 & 34.78 \\
\phantom{0}3 & 6.92 & 0.39 & 1.55 & 43.64 \\
\phantom{0}4 & 4.68 & 0.76 & 1.05 & 50.13 \\
\phantom{0}5 & 3.02 & 0.52 & 0.80 & 54.46 \\
\phantom{0}6 & 2.44 & 0.29 & 0.69 & 57.88 \\
\phantom{0}7 & 1.63 & 0.77 & 0.54 & 60.82 \\
\phantom{0}8 & 1.48 & 0.67 & 0.39 & 63.36 \\
\phantom{0}9 & 1.41 & 0.51 & 0.35 & 65.64 \\
10 & 1.19 & 0.38 & 0.33 & 67.54 \\[3pt]
& 41.09 & 13.17 & 13.28 & 67.54 \\
\hline
\end{tabular*}
\end{table}

%= 0.35]{model_1_msty_W_2}
%= 0.35]{model_1_edss_W_2}
%

%f9 #&#
\begin{figure}

\includegraphics{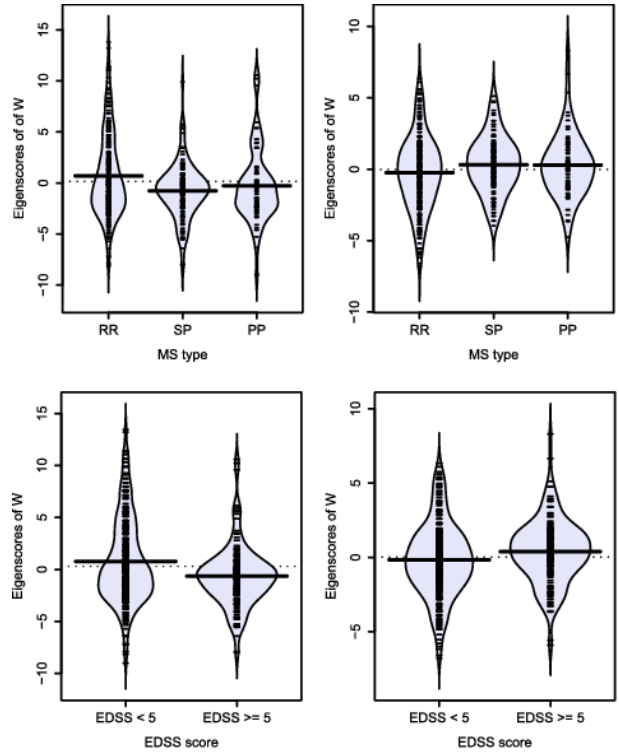}

\caption{Model 2: Group beanplots according to MS type (top) and
according to EDSS score (bottom).}
\label{figure:model_2_W}
\end{figure}

%= 0.35]{model_2_msty_W_2}
%= 0.35]{model_2_edss_W_2}
%

}

%s6 #&#
\section{Discussion}\label{sec:Discussion}
The methods developed in this paper increase the scope and general
applicability of LFPCA to very high-dimensional settings. The base
model decomposes the longitudinal data into three main components: a
subject-specific random intercept, a subject-specific random slope, and
reversible visit-to-visit deviation. We described and addressed
computational difficulties that arise with high-dimensional data using
a powerful approach referred to as HD-LFPCA. We have developed a
procedure designed to identify a low-dimensional space that contains
all the information for estimating of the model. This significantly
extended the previous related efforts in the clustered functional
principal components models, MFPCA [\citet{Chong-Zhi:etal:2008}] and
HD-MFPCA [\citet{Zipunnikov:etal:2011a}].

We applied HD-LFPCA to a novel imaging setting considering DTI and MS
in a primary white matter structure. Our investigation characterized
longitudinal and cross-sectional variation in the corpus callosum.

There are several outstanding issues for HD-LFPCA that need to be
addressed. First, a key assumption of our methods is that they require
a moderate sample size that does not exceed ten thousands, or so,
images. This limitation can be circumvented by adopting the methods
discussed in the \hyperref[app]{Appendix}. Second, we have not formally included white
noise in our model. Simulation studies in Section~\ref{sec:Simulations}
demonstrated that a moderate amount of white noise does not have a
serious effect on the estimation procedure. However, a more systematic
treatment of the related issues is required.

In summary, HD-LFPCA provides a powerful conceptual and practical step
toward developing estimation methods for structured
ultrahigh-dimensional data.

%summarizing the simulation studies. It also contains additional
%results of the analysis of DTI images of MS patients.}

\begin{appendix}\label{app}
\section*{Appendix}

\subsection{Large sample size}\label{subsec:large-sample-size}
The main assumption which has been made in the paper is that the sample
size, $n=\sum_{j=1}^IJ_i$, is sufficiently small to guarantee that
calculations of order $O(n^3)$ are feasible. Below we briefly describe
how our framework can be adapted to settings with many more scans---on
the order of tens or hundreds of thousands.

LFPCA equation (\ref{eq:01}) models each vector $\bbY_{ij}$ as a linear
combination of columns of matrices $\bPhi^{X,0}$, $\bPhi^{X,1}$,
$\bPhi
^W$. {Assuming that $2N_X + N_W < n$, each $\bbY_{ij}$ belongs
to an at most $(2N_X+N_W)$-dimensional linear space $\mathfrak
{L}(\bPhi
^{X,0}, \bPhi^{X,1}, \bPhi^{W})$ spanned by those columns.} Thus, if
model (\ref{eq:01}) holds exactly the rank of the matrix, $\bbY$ does
not exceed $(2N_X+N_W)$ and at most $2N_X+N_W$ columns of $\bV$
correspond to nonzero singular values. This implies that the intrinsic
model (\ref{eq:04}) can be obtained by projecting onto the first
$2N_X+N_W$ columns of $\bV$ and the sizes of matrices $\bA^{X,0},\bA
^{X,1},\bA^W$ in (\ref{eq:04}) will be $(2N_X+N_W)\times N_X$,
$(2N_X+N_W)\times N_X$, and $(2N_X+N_W)\times N_W$, respectively.
Therefore, the most computationally intensive part would require
finding the first $2N_X+N_W$ left singular vectors of $\bbY$. Of
course, in practice, model (\ref{eq:01}) never holds exactly. Hence,
the number of columns of matrix $\bV$ should be chosen to be large
enough to either reasonably exceed $(2N_X+N_W)$ or to capture the most
variability in data. The latter can be estimated by tracking down the
sums of the squares of the corresponding first singular vectors. Thus,
this provides a constructive way to handle situations when $n$ is too
large to calculate the SVD of~$\bbY$.

There are computationally efficient ways to calculate the first $k$
singular vectors of a large matrix. One way is to adapt streaming
algorithms [\citet{Weng:etal:2003,Zhao:etal:2006,Budavari:2009}].
These algorithms usually require only one pass through the data matrix
$\bbY$ during which information about the first $k$ singular vectors is
accumulated sequentially. Their complexity is of order $O(k^3p)$. An
alternate approach is to use iterative power methods [see, e.g.,
\citet{Roweis:1997}]. As the dimension of the intrinsic model,
$2N_X+N_W$, is not known in advance, the number of left singular
vectors to keep and project\vspace*{1pt} onto can be adaptively estimated based on
the singular values of the matrix $\bbY$. Further development in this
direction is beyond the scope of this paper.

\subsection{Proofs}\label{subsec:MM-quadratics}
\mbox{}
\begin{pf*}{Proof of Lemma~\ref{le1}}
Using the independence of $\bY_i$ and $\bY_k$, the expectation
of pairwise quadratics is
%
%e13 #&#
\begin{eqnarray}
\label{eq:MMQ-01} &&E\bigl(\bY_{ij_1}\bY_{kj_2}^{\prime}\bigr)
\nonumber
\\[-8pt]
\\[-8pt]
\nonumber
&&\qquad= \cases{ %
{\bolds\eta} {\bolds\eta}^{\prime},
\qquad\mbox{if } k \neq i,
\vspace*{2pt}\cr
{\bolds{\eta}} {\bolds{\eta}}^{\prime} +\bK_X^{00}+T_{ij_2}
\bK _X^{01}+T_{ij_1}\bK_X^{10}+
T_{ij_1}T_{ij_2} \bK_X^{11}+\delta
_{j_1j_2}\bK^W,\vspace*{2pt}\cr
\hspace*{42pt}\mbox{if } i = k,}
\end{eqnarray}
where $\delta_{j_1j_2}$ is $1$ if $j_1=j_2$ and $0$ otherwise. From the
top equality we get the MM estimator of the mean, $\hbeta= n^{-1}\sum_{i,j}\bY_{ij}$. The covariances $\bK^X$ and $\bK^W$ can be estimated
by de-meaning $\bY_{ij}$ as $\bbY_{ij} = \bY_{ij}-\hbeta$ and
regressing $\bbY_{ij_1}\bbY_{ij_2}^{\prime}$ on $1, T_{ij_2}, T_{ij_1},
T_{ij_1}T_{ij_2}$, and $\delta_{j_1j_2}$. The bottom equality can be
written as $E(\bbY_{ij_1j_2}^v) = \bK^v \bff_{ij_1j_2}$,
%E\bbY_{ij_1j_2}^v = \bK^v \bff_{ij_1j_2},
where $\bbY_{ij_1j_2}^v = \bbY_{ij_2}\otimes\bbY_{ij_1}$ is a
$p^2\times1$ dimensional vector, the parameter of interest is the
$p^2\times5$ matrix $\bK^v = [\operatorname{vec}(\bK_X^{00}),
\operatorname{vec}(\bK_X^{01}), \operatorname{vec}(\bK_X^{10}),
\operatorname{vec}(\bK_X^{11})$,
$ \operatorname{vec}(\bK^W)]$, and the covariates are entries in the
$5\times
1$ vector\vspace*{1pt} $\bff_{ij_1j_2}=(1,T_{ij_2}, T_{ij_1},T_{ij_1}T_{ij_2},\delta
_{j_1j_2})^{\prime}$. With this notation $E\bY^v = \bK^v \bF$,
where $\bbY^v$ is $p^2\times m$ dimensional with $m = \sum_{i=1}^IJ_i^2$ and $\bF$ is a $5\times m$ dimensional matrix with
columns equal to $\bff_{ij_1j_2},i=1,\ldots,I$ and $j_1,j_2 =
1,\ldots,
J_i$. The MM estimator of $\bK^v$ is thus $\bhK^v =\bbY^v\bF
^{\prime}(\bF\bF
^{\prime})^{-1}$,
which provides unbiased estimators of the covariances $\bK^X$ and $\bK
^W$. If we denote $\bH= \bF^{\prime}(\bF\bF^{\prime})^{-1}$, we get the result
of the lemma.
\end{pf*}

\begin{pf*}{Proof of Lemma~\ref{le2}} Let us denote by $\hat{\bK}_{\bU}^X$ and
$\hat
{\bK}_{\bU}^W$ the matrices defined by equations (\ref{eq:OLS-03}) with
$\bS^{1/2}\bU_{ij_1}\bU_{ij_2}^{\prime}\bS^{1/2}$ substituted for $\bbY
_{ij_1}\bbY_{ij_2}^{\prime}$. The $2n\times2n$ dimensional matrix $\hat
{\bK
}_{\bU}^X$ and the $n\times n$ dimensional matrix $\hat{\bK}_{\bU}^W$
are low-dimensional counterparts of $\hat{\bK}^X$ and $\hat{\bK}^W$,
respectively. Using the SVD representation $\bbY_{ij} = \bV\bS
^{1/2}\bU
_{ij}$, the estimated high-dimensional covariance matrices can be
represented as $\hat{\bK}^X=\bD\hat{\bK}_{\bU}^X\bD'$ and $\hat
{\bK
}^W=\bV\hat{\bK}_{\bU}^W\bV'$, where the matrix $\bD$ is
$2p\times2n$
dimensional with orthonormal columns defined as
%
%e14 #&#
\begin{equation}
\bD= %
\pmatrix{\bV& \bZero_{p\times n} \vspace*{2pt}
\cr
\bZero_{p\times n} & \bV } %
.
\end{equation}
From the constructive definition of $\bH$, it follows that the matrices
$\hat{\bK}_{\bU}^X$ and $\hat{\bK}_{\bU}^W$ are symmetric. Thus,
we can
construct their spectral decompositions, $\hat{\bK}_{\bU}^X =\bhA
^{X}\hat{\bLambda}^X\bhA^{X'}$ and $\hat{\bK}_{\bU}^W = \bhA
^{W}\hat
{\bLambda}^W\bhA^{W'}$. Hence, high-dimensional covariance matrices can
be represented as $\hat{\bK}^X=\bD\bhA^{X}\hat{\bLambda}^X\bhA
^{X'}\bD
'$ and $\hat{\bK}^W=\bV\bhA^{W}\hat{\bLambda}^W\bhA^{W'}\bV'$,
respectively. The result of the lemma now follows from the
orthonormality of the columns of matrices $\bD$ and $\bV$.
\end{pf*}

\begin{pf*}{Proof of Lemma~\ref{le3}} With notational changes, the proof is identical
to the proof of Lemma~\ref{le1}.
\end{pf*}
\begin{pf*}{Proof of Lemma~\ref{le4}} With notational changes, the proof is identical
to the proof of Lemma~\ref{le2}.
\end{pf*}
\begin{pf*}{Proof of Lemma~\ref{le5}} The main idea of the proof is similar
to that of \citet{Zipunnikov:etal:2011a}. We assume that function
$\eta
(v,T_{ij})=0$. From the model it follows that $\bomega_i \sim
(0,\bLambda_{\bomega})$, where
$\bLambda_{\bomega}$ is a covariance matrix of $\bomega_i$.
When $p \le N_X+J_iN_W$ the BLUP of $\bomega_i$ is given by $\hat
{\bomega}_i = \operatorname{Cov}(\bomega_i,\operatorname{vec}(\bbY_i))\times\break \operatorname{Var}(\operatorname{vec}(\bbY
_i))^{-1} \operatorname{vec}(\bbY
_i) = \bLambda_{\bomega}\bB_i^{\prime}(\bB_i\bLambda_{\bomega}\bB
_i^{\prime})^{-1}\operatorname{vec}(\bbY_i)$ [see McCulloch and\break 
Searle (\citeyear{McCulloch:Searle:2001}), Section~9].
The BLUP is essentially a projection and, thus, it does not require any
distributional assumptions. It may be defined in terms of a projection
matrix. If $\bxi_i$ and $\bzeta_{ij}$ are normal, then the BLUP is the
best predictor. When $p > N_X+J_iN_W$ the matrix $\bB_i\bLambda
_{\bomega
}\bB_i^{\prime}$ is not invertible and the generalized inverse of $\bB
_i\bLambda_{\bomega}\bB_i^{\prime}$ is used [\citet{Harville:1976}]. In that
case, $\hat{\bomega}_i = \bLambda_{\bomega}\bB_i^{\prime}(\bB
_i\bLambda
_{\bomega}\bB_i^{\prime})^{-}\operatorname{vec}(\bbY_i)=
\bLambda_{\bomega}^{1/2}(\bLambda_{\bomega}^{1/2}\bB_i^{\prime}\bB
_i\bLambda
_{\bomega}^{1/2})^{-1}\bLambda_{\bomega}^{1/2}\bB_i^{\prime}\operatorname{vec}(\bbY_i)
=(\bB
_i^{\prime}\bB_i)^{-1}\bB_i^{\prime}\operatorname{vec}(\bbY_i)$. Note that it coincides with the
OLS estimator for $\bomega_i$ if $\bomega_i$ were a fixed parameter.
Thus, the estimated BLUPs are given by
$\hat{\bomega}_i =(\hat{\bB}_i^{\prime}\hat{\bB}_i)^{-1}\hat{\bB
}_i^{\prime}\operatorname{vec}(\bbY_i)$.
\end{pf*}
\end{appendix}

\section*{Acknowledgments}
The authors would like to thank Jeff Goldsmith for his help with data
management.
%The research of Vadim Zipunnikov, Brian Caffo, and Ciprian
%Crainiceanu was supported by grant R01NS060910 from the National
%Institute of Neurological Disorders and Stroke and by Award Number
%EB012547 from the NIH National Institute of Biomedical Imaging and
%Bioengineering (NIBIB). The research of Sonja Greven was funded by the
%German Research Foundation through the Emmy Noether Programme, grant GR
%3793/1-1. The research of Daniel S. Reich was supported by the
%Intramural Research Program of the National Institute of Neurological
%Disorders and Stroke.
The content is solely the responsibility of the
authors and does not necessarily represent the official views of the
National Institute of Neurological Disorders and Stroke or the National
Institute of Biomedical Imaging and Bioengineering or the National
Institutes of Health.

\begin{supplement}[id=suppA]
\stitle{Supplement to ``Longitudinal high-dimensional
principal components analysis with application
to diffusion tensor imaging of multiple sclerosis''}
\slink[doi]{10.1214/14-AOAS748SUPP} %[doi,text={...}] - jei
%reikiasuskaldyti doi
\sdatatype{.pdf}
\sfilename{aoas748\_supp.pdf}
\sdescription{We provide extra figures and tables summarizing the results
of simulation studies and the analysis of DTI images of MS patients.}
\end{supplement}

% imsref loaded by akundreckaite, 2014-09-23 15:28:37
% imsref loaded by akundreckaite, 2014-09-23 15:42:20
%

% zodis "Acknowledgments" paliekamas pagal autoriu

%suskaldyti doi

\printaddresses
\end{document}